\documentclass[11pt]{article}

\usepackage{verbatim}
\usepackage{amsmath}
\usepackage{amsfonts}
\usepackage{amssymb}
\usepackage{bbm}
\usepackage{latexsym}
\usepackage{lscape} 
\usepackage{epsfig}
\usepackage{color}
\usepackage{graphicx}
\usepackage[table]{xcolor}
\usepackage{enumerate}
\usepackage{hhline}
\usepackage{subfig}
\usepackage{multirow}
\usepackage{mathrsfs}

\usepackage{amscd}
\usepackage{cite}

\DeclareMathSymbol{\mg}{\mathrel}{symbols}{"1D}

\newcommand{\gL}{\Lambda}

\newcommand{\tr}{\text{tr}}

\newcommand{\ra}{\rightarrow}

\newcommand{\dsp}{\displaystyle}
\newcommand{\scp}{\scriptstyle}

\newcommand{\beq}{\begin{equation}}
\newcommand{\eeq}{\end{equation}}
\newcommand{\barr}{\begin{array}}
\newcommand{\earr}{\end{array}}

\newcounter{oldcounter}

\newcommand{\Intr}{\mathbb{Z}}

\newcommand{\Real}{\mathbb{R}}

\newcommand{\ba}[2]{\[\begin{array}{#2}\label{#1}}
\newcommand{\ea}{\end{array}\]}
\newcommand{\be}{\begin{equation}}
\newcommand{\ee}{\end{equation}}
\newcommand{\bea}{\begin{eqnarray}}
\newcommand{\eea}{\end{eqnarray}}

\newcommand{\brkt}[2]{\left[{\scp #1}\atop{\scp #2}\right]}

\newcommand{\rep}[1]{\mathbf{#1}}

\newcommand{\brep}[1]{\overline{\rep{#1}}}

\newcommand{\sm}{{\,\mbox{-}}}

\begin{document}

\thispagestyle{empty}

\begin{flushright}
LMU-ASC 47/14 \\
MITP/14-052
\\
\end{flushright}
\begin{center}
{\Large {\bf 
Non-supersymmetric heterotic model building 
} 
}
\\[0pt]

\bigskip
\bigskip {\large
{\bf 
{Michael Blaszczyk}$^{a,}$}\footnote{
E-mail: blaszcz@uni-mainz.de},
{\bf 
{Stefan Groot Nibbelink}$^{b,}$}\footnote{
E-mail: Groot.Nibbelink@physik.uni-muenchen.de},
{\bf 
{Orestis Loukas}$^{b,c,}$}\footnote{
E-mail: O.Loukas@physik.uni-muenchen.de}, 
\\[1ex]  
{\bf 
{Sa\'ul Ramos-S\'anchez}$^{d,}$}\footnote{
E-mail: ramos@fisica.unam.mx}
\bigskip }\\[0pt]
\vspace{0.23cm}
${}^a$ {\it 
PRISMA Cluster of Excellence \& Institut f\"ur Physik (WA THEP), \\ 
Johannes-Gutenberg-Universit\"at, 55099 Mainz, Germany
 \\[1ex] } 
${}^b$ {\it 
Arnold Sommerfeld Center for Theoretical Physics,   \\ 
Ludwig-Maximilians-Universit\"at M\"unchen, 80333 M\"unchen, Germany
}
\\[1ex] 
${}^c$ {\it 
Physics Department, National Technical University of Athens, \\ 
Zografou Campus, 15780 Athens, Greece
}\\[1ex] 
${}^d$ {\it 
Department of Theoretical Physics, Physics Institute, UNAM \\
Mexico D.F. 04510, Mexico
}
\\[1ex] 
\bigskip
\end{center}

\subsection*{\centering Abstract}

We investigate orbifold and smooth Calabi-Yau compactifications of the non-supersymmetric heterotic SO(16)$\times$SO(16) string. 
We focus on such Calabi-Yau backgrounds in order to recycle commonly employed techniques, like index theorems and cohomology theory, 
to determine both the fermionic and bosonic 4D spectra. 
We argue that the N=0 theory never leads to tachyons on smooth Calabi-Yaus in the large volume approximation.
As twisted tachyons may arise on certain singular orbifolds, we conjecture that such tachyonic states are lifted in the full blow-up. 
We perform model searches on selected orbifold geometries. 
In particular, we construct an explicit example of a Standard Model-like theory with three generations and a single Higgs field.

\newpage 
\setcounter{page}{1}
 \setcounter{footnote}{0}
\tableofcontents
\newpage

\section{Introduction}
\label{sc:Introduction}


The central emphasis in the heterotic string phenomenology community during the last 20 years or so has been on the construction of supersymmetric Standard Model(MSSM)-like models from string theory. 
Heterotic model building on smooth Calabi--Yau spaces with non--Abelian vector bundles~\cite{Candelas:1985en} has resulted in MSSM--like models~\cite{Bouchard:2005ag} with possible supersymmetry breaking built in~\cite{Braun:2005ux,Braun:2005bw,Braun:2005nv}. More recently, similar models have been constructed using line bundles instead~\cite{Anderson:2011ns,Anderson:2012yf,Anderson:2013xka}. 
Heterotic orbifolds~\cite{Dixon:1985jw,Dixon:1986jc,Ibanez:1986tp,Ibanez:1987pj} may also be used to construct MSSM--like models, see e.g.~\cite{string_compactification_phys_rept,Choi2006}. 
In Refs.~\cite{Buchmuller:2005jr,Buchmuller:2006ik,Lebedev:2006kn,Lebedev:2008un} MSSM--like models have been assembled on the toroidal ${\mathbb Z}_\text{6--II}$ orbifold. 
Similar searches have been performed on the $\Intr_2\times \Intr_4$ orbifold~\cite{Z2xZ4}, $\Intr_\text{12--I}$ orbifold~\cite{Kim:2006hv,Kim:2007mt} and $\Intr_8$ orbifolds~\cite{Nibbelink:2013lua}. For a comprehensive overview of model building on various orbifold geometries see~\cite{Nilles:2014owa}. 
Essentially all these orbifold models break the E$_8\times$E$_8$ gauge group of the heterotic string directly down to the SM gauge group. 
To avoid that the hypercharge will be broken if such models are fully resolved~\cite{Nibbelink:2009sp}, one may break the SU(5) Grand Unified (GUT) gauge group by a non--local freely acting Wilson line~\cite{Blaszczyk:2009in,Blaszczyk:2010db}.


In this work we investigate the construction of non-supersymmetric models in string theory. 
The main motivation for this work is that so far the LHC or any other experiment has not found any hint for the existence of supersymmetry in particle physics. 
Currently, the bound on the supersymmetry breaking is of the order of 1 TeV~\cite{Beringer:1900zz}. 
This means that one of the initial motivations for supersymmetry, namely to solve the hierarchy problem,  becomes less convincing as one is still left with a sizable hierarchy between the Higgs mass and the supersymmetry breaking scale. 
Moreover, the MSSM has over 120 free parameters most of which are associated with soft-supersymmetry breaking masses and couplings. 
The present work originated from the following questions\footnote{Inspired by a discussion with Brent Nelson at ICTP.}: Suppose that the world is not supersymmetric up to the Planck scale or beyond, can we still use string theory as a framework to study particle physics? 
If so, how close could we get to the Standard Model? 
In this scenario, we take the most extreme point of view: 
We assume that target space supersymmetry does not even exist at the Planck / string scale.


When one considers non-supersymmetric models from string theory there are various potentially problematic issues. 
The most prominent ones are the following: 
\begin{enumerate}[i.)]
\item
The spectrum might contain level-matched tachyons.
\item 
One loses practical computational control since one now has to study compactifications on generic six dimensional internal manifolds. 
\item 
The Higgs mass will be quadratically dependent on the high scale. 
\item 
In general one expects a cosmological constant of the order of the string scale. 
\item 
Associated with the cosmological constant, a destabilizing dilaton tadpole is generated.  
\end{enumerate}
In this work we will only aim to investigate the first two problems in detail. 
Even though the latter issues are very serious, they are only considered in the outlook
at the end of this work, where the main hurdles and some previous efforts to tackle them
are discussed. It is nonetheless important to stress that solving them is a crucial
challenge not only for non-supersymmetric string constructions, but also for field-theoretical
descriptions of physics beyond the Standard Model, where e.g.\ the cosmological constant
even including supersymmetry requires more compelling explanations.


In the past there have been various works addressing non-supersymmetric models from string theory. 
Dienes~\cite{Dienes:1994np,Dienes:2006ut} performs some statistical scan of non-supersymmetric free-fermion models~\cite{Shiu:1998he} to give some idea of the scattering of the value of the cosmological constant. 
The connection between non-supersymmetric free fermionic models~\cite{Kawai:1986vd}, the Ho\v{r}ava-Witten model and other dualities have been studied in~\cite{Blum:1997cs,Faraggi:2007tj}.  
A large set of non-supersymmetric models in four dimensions were constructed using a covariant lattice approach~\cite{Lerche:1986ae,Lerche:1986cx}.
A strategy based on the inclusion of NS5-branes was applied to obtain a class of non-supersymmetric heterotic flux vacua with torsion~\cite{Held:2010az}.
Finally, non-supersymmetric tachyon-free type-I/II orientifold models 
\cite{Sagnotti:1995ga,Sagnotti:1996qj,Angelantonj:1998gj,Sugimoto:1999tx,
Blumenhagen:1999ns, Aldazabal:1999tw,Moriyama:2001ge}
have also been constructed as rational conformal field theories~\cite{GatoRivera:2007yi,GatoRivera:2008zn}. 


In order to better understand the four-dimensional non-supersymmetric theories emerging from string theory, we take as our starting point the ten dimensional non-supersymmetric SO(16)$\times$SO(16) theory~\cite{Dixon:1986iz,Dixon:1986jc,AlvarezGaume:1986jb}; its low energy spectrum is given in Table~\ref{tb:N=0Spectrum}. 
We use two formulations of this theory: 
One which is inspired by a fermionic description with non-trivial torsion phases between the three spin structures, while the other uses a $\Intr_2$ orbifolding of a lattice formulation of the heterotic string. 
Narain compactification~\cite{narain_86,narain_86_2} of this theory has been considered in Ref.~\cite{Nair:1986zn,Ginsparg:1986wr}. 
More general compactifications of this theory were investigated in Ref.~\cite{Lust:1986kj} implementing the Fischler-Susskind mechanism~\cite{Fischler:1986ci,Fischler:1986tb}. 
In this work we compactify this non-supersymmetric theory on {\it Calabi-Yau spaces}, 
such as orbifolds~\cite{Dixon:1985jw} and smooth Calabi-Yau manifolds~\cite{Candelas:1985en} with (a discrete subgroup of) SU(3) holonomy. 
To the best of our knowledge, 
on non-supersymmetric (a)symmetric orbifolds have been considered in the past in Ref.~\cite{Taylor:1987uv,Toon:1990ij,Sasada:1995wq,Font:2002pq}.

Given that we are investigating non-supersymmetric models, there seems to be no need to consider string backgrounds that would preserve some supersymmetry in principle. 
However, we think it is advantageous to nevertheless consider -- would be -- supersymmetry-preserving geometries for various reasons.
Firstly, in the fermionic formulation without further torsion phases mixing the orbifold and spin structures, one finds the restriction to supersymmetric twists.
Secondly, we shall be able to recycle many of the techniques developed for smooth Calabi-Yau manifolds, to compute the four dimensional spectra of both fermions and bosons obtained from the compactification of the non-supersymmetric theory. 
Finally, even just using the standard embedding we find an SO(10) GUT theory with fermions living in the spinor $\mathbf{16}$ representation and scalars in the $\mathbf{10}$, which may be considered as a sign of the phenomenological potential of these string constructions.

\begin{table} 
\begin{center} 
  \renewcommand{\arraystretch}{1.2}
\begin{tabular}{|c | l |}
\hline 
Fields & Space-time interpretation 
\\ \hline\hline 
 \cellcolor{lightgray} $G_{MN}, B_{MN}, \phi$ 
 &  \cellcolor{lightgray}Graviton, Kalb-Ramond 2-form, Dilaton
\\
 \cellcolor{lightgray}$A_M$ 
 &  \cellcolor{lightgray} SO(16)$\times$SO(16) Gauge fields 
\\ \hline 
$\Psi_+$ & Spinors in the $(\mathbf{128},\mathbf{1})+(\mathbf{1},
\mathbf{128})$ 
\\
$\Psi_-$ & Cospinors in the $(\mathbf{16},\mathbf{16})$
\\ \hline 
\end{tabular}
  \renewcommand{\arraystretch}{1}
\end{center}
\caption{\label{tb:N=0Spectrum}
Massless spectrum of the ten-dimensional N=0, SO(16)$\times$SO(16) heterotic theory. Bosons and fermions are indicated with gray and white background, respectively, in this and most subsequent tables.}
\end{table}

\subsection*{Paper overview}

In Section~\ref{sc:10DN=0} we review two descriptions of the non-supersymmetric heterotic SO(16)$\times$SO(16) string. 
Next, we consider compactification of this theory on orbifolds that would themselves preserve N=1 supersymmetry in Section~\ref{sc:Orbifold}. 
We  illustrate such orbifold constructions by some specific tachyon-free $\Intr_3$ orbifold models and discuss a tachyonic $\Intr_\text{6-I}$ model. 
In Section~\ref{sc:Smooth} we exploit that our backgrounds still preserve supersymmetry to compute the spectra of both fermionic and bosonic zero modes.
We argue that no tachyons arise on smooth Calabi-Yau manifolds and illustrate this fact by considering the blow-up of a tachyonic $\Intr_\text{6-I}$ model. 
Section~\ref{sc:Models} is devoted to model scans and searches for non-supersymmetric Standard Model-like models. 
Finally, in Section~\ref{sc:Conclusion} we recapitulate our main findings and give an outlook on open issues. 
Appendix~\ref{sc:ThetaLattice} gives details of the partition functions and lattices used in the text.

\section{Non-supersymmetric heterotic string}
\label{sc:10DN=0}

\subsection{The standard N=1 supersymmetric E$_8\times$E$_8$ string}

The N=1 supersymmetric 10D heterotic E$_8\times$E$_8$ string theory~\cite{Gross:1985fr,Gross:1985rr} has three $\mathbb{Z}_2$ twists
\begin{equation}
\psi^i(\sigma+1) = e^{2\pi i \frac s2} \, \psi^i(\sigma)~, 
\quad 
\lambda^I_1(\sigma+1) = e^{2\pi i \frac {t}2}\, \lambda^I_1(\sigma)~, 
\quad 
\lambda^I_2(\sigma+1) = e^{2\pi i \frac {u}2}\, \lambda^I_2(\sigma)~, 
\label{eq:coundaryconditions1}
\end{equation}
in the fermionic construction. Here $\psi^i$, $i=0,1,2,3$, are complex right-moving fermions and $\lambda^I_{1,2}$, $I=1,\ldots 8$, two sets of complex left-moving fermions. At the one-loop level we have similar boundary conditions for $\sigma \ra \sigma-\tau$ which we label by $s',t',u'$. In total this leads to $2^{2\cdot3} = 64$ terms in the one-loop partition function 
\begin{equation} 
{\mathbf Z}_{\text{E}_8^2}(\tau,\overline\tau) = 
- 
\frac18 \sum
{\mathbf Z}_8^x(\tau,\overline\tau) \cdot 
\widehat {\mathbf Z}_4\brkt{\frac s2  e_4}{\frac {s'}2  e_4}(\tau) \cdot 
\overline{ \widehat {\mathbf Z}_8\brkt{\frac t2  e_8}{\frac {t'}2  e_8 }(\tau)} \cdot 
\overline{ \widehat {\mathbf Z}_8\brkt{\frac u2  e_8}{\frac {u'}2 e_8 }(\tau)}~,
\end{equation}
where the sum is over all spin structures $s,s',t,t',u,u'=0,1$. 
(The sign in front of this expression ensures that target space bosons / fermions give a positive / negative contribution to the full partition function.) 
The bosonic and fermionic partition functions, ${\mathbf Z}_D^x$ and $\widehat {\mathbf Z}_d$, are defined in \eqref{sector_phase} and \eqref{free_boson_partition} of Appendix~\ref{sc:ThetaLattice}. This partition function is modular invariant and leads to a tachyon-free spectrum. At the massless level one finds the well-known spectrum of N=1 supergravity  coupled to E$_8\times$E$_8$ super Yang-Mills theory in ten dimensions.

\subsection{Non-supersymmetric generalized torsion phases}

Modular invariance and absence of tachyons do not fix the theory with three $\mathbb{Z}_2$ twists uniquely~\cite{AlvarezGaume:1986jb}: One can introduce various modular invariant phases in the partition function above. Most of them lead to trivial flippings of the chiralities of the various spinor representations. However, up to such equivalences there is one further modular invariant and tachyon-free partition function:
\begin{equation} \label{SO(16)_partition}
{\mathbf Z}_\text{ferm}(\tau,\overline\tau) = 
- 
\frac18 \sum
T \cdot 
{\mathbf Z}_8^x(\tau,\overline\tau) \cdot 
\widehat {\mathbf Z}_4\left[^{\frac s2  e_4}_{\frac {s'}2 e_4 }\right](\tau) \cdot 
\overline{ \widehat {\mathbf Z}_8\brkt{\frac t2  e_8}{\frac {t'}2 e_8 }(\tau)} \cdot 
\overline{ \widehat {\mathbf Z}_8\brkt{\frac u2  e_8}{\frac {u'}2 e_8 }(\tau)}~,
\end{equation}
obtained from the E$_8\times$E$_8$ partition function by introducing a generalized torsion phase $T_\text{torsion}$ given by $T = T_\text{torsion} \cdot T_\text{chiral}$ with 
\begin{equation} \label{TorsionPhases} 
T_\text{torsion} = 
(-)^{s t' - s' t} (-)^{s u' - s' u}  (-)^{t u' - t' u}~, 
\qquad 
T_\text{chiral} = 
- (-)^{s's+s'+s} (-)^{t't + t' +t} (-)^{u'u + u' + u }
 ~. 
\end{equation} 
In addition, we have introduced the phase $T_\text{chiral}$ which simply interchanges the spinor with cospinor lattices for later convenience.

The partition function~\eqref{SO(16)_partition} encodes a fundamentally different theory from the supersymmetric E$_8\times$E$_8$ theory. In particular, it is not supersymmetric as the massless spectrum, given in Table~\ref{tb:N=0Spectrum}, clearly shows: 
It contains the bosonic states of the supergravity multiplet, but not its gravitino and dilatino. The gauge group is SO(16)$\times$SO(16) rather than E$_8\times$E$_8$, but without gauginos corresponding to this gauge group. Instead, we encounter ten dimensional chiral fermions in the spinor representations of both of these SO(16) gauge groups, and a chiral fermion of the opposite chirality in the bi-fundamental representation of both.

Supersymmetry is neither present at the massless level nor in the full tower of string excitations. To see this very clearly it is instructive to write the full fermionic partition function in a lattice formulation. For each of the eight sectors ($s,t,u =0,1$) one can determine to which lattice it is associated. In Table~\ref{tb:Lattices_E8_N=0} we have listed the lattices for these eight sectors for both the supersymmetric E$_8\times$E$_8$ theory and the non-supersymmetric SO(16)$\times$SO(16) theory. Here we see a couple of crucial differences between the two theories: 
In the supersymmetric theory only a restricted set of lattices appears. 
Concretely, on the right-moving side we either encounter the spinor {\bf S}$_4$ or vector{\bf V}$_4$ lattices, while on the left-moving side only the SO(16) root {\bf R}$_8$ and spinor {\bf S}$_8$ lattices. In the non-supersymmetric theory all four possible lattices appear on both the left- and right-moving side. Moreover, the direct sum of all eight lattices in the supersymmetric case can be factorized as 
$\boldsymbol{\Gamma}_4 \otimes \text{\bf E}_8 \otimes \text{\bf E}_8$ (using the definitions in Table~\ref{tb:Lattices}) which reflects at all mass levels that the theory is supersymmetric and has the E$_8\times$E$_8$ symmetry structure. Clearly, such a factorization is impossible for the non-supersymmetric theory.

\begin{table}
\begin{center}
  \renewcommand{\arraystretch}{1.2}
 \begin{tabular}{|c||c|c|c|c|c|}
 \hline
  Sector   & \multicolumn{2}{|c|}{Lattices in the theory}   \\
 (s,t,u)  &  ~~~~N=1, E$_8\times$E$_8$~~~~ & N=0, SO(16)$\times$SO(16)
 \\\hline\hline 
 (1,1,1)  & {\bf V}$_4$ $\otimes$ {\bf R}$_8$ $\otimes$ {\bf R}$_8$ &  {\bf V}$_4$ $\otimes$ {\bf R}$_8$ $\otimes$ {\bf R}$_8$  
 \\\hline
 (1,0,0)  & {\bf V}$_4$ $\otimes$ {\bf S}$_8$\, $\otimes$ {\bf S}$_8$\, &  {\bf V}$_4$ $\otimes$ {\bf S}$_8$\, $\otimes$ {\bf S}$_8$\,   
 \\\hline
 \cellcolor{lightgray}(1,0,1)  &  \cellcolor{lightgray}{\bf V}$_4$ $\otimes$ {\bf S}$_8$\, $\otimes$ {\bf R}$_8$ &  \cellcolor{lightgray}{\bf R}$_4$ $\otimes$ {\bf C}$_8$ $\otimes$ {\bf V}$_8$  
 \\\hline
  \cellcolor{lightgray}(1,1,0)  &  \cellcolor{lightgray}{\bf V}$_4$ $\otimes$ {\bf R}$_8$ $\otimes$ {\bf S}$_8$\, &   \cellcolor{lightgray}{\bf R}$_4$ $\otimes$ {\bf V}$_8$ $\otimes$ {\bf C}$_8$  
 \\\hline\hline 
  (0,0,1)  & {\bf S}$_4$\, $\otimes$ {\bf S}$_8$\, $\otimes $ {\bf R}$_8$ &  {\bf S}$_4$\, $\otimes$ {\bf S}$_8$\, $\otimes$ {\bf R}$_8$  
 \\\hline
 (0,1,0)  & {\bf S}$_4$\, $\otimes$ {\bf R}$_8$ $\otimes $ {\bf S}$_8$\, &  {\bf S}$_4$\, $\otimes$ {\bf R}$_8$ $\otimes$ {\bf S}$_8$  
 \\\hline
 \cellcolor{lightgray}(0,1,1)  &  \cellcolor{lightgray}{\bf S}$_4$ $\otimes$ {\bf R}$_8$ $\otimes$ {\bf R}$_8$ &   \cellcolor{lightgray}{\bf C}$_4$ $\otimes$ {\bf V}$_8$ $\otimes$ {\bf V}$_8$  
 \\\hline
 \cellcolor{lightgray}(0,0,0)  &  \cellcolor{lightgray}{\bf S}$_4$ $\otimes$ {\bf S}$_8$\, $\otimes$ {\bf S}$_8$\, &  \cellcolor{lightgray}{\bf C}$_4$ $\otimes$ {\bf C}$_8$ $\otimes$ {\bf C}$_8$  
 \\\hline
 \end{tabular}
   \renewcommand{\arraystretch}{1}
\end{center}
\caption{\label{tb:Lattices_E8_N=0}
The different lattices that occur in the eight different sectors of the supersymmetric E$_8\times$E$_8$ and the non-symmetric SO(16)$\times$SO(16) heterotic string theories. 
The white / gray background entries in the last column correspond to the untwisted / twisted sectors of the supersymmetry breaking twist.
(The definition of the lattices can be found in Table~\ref{tb:Lattices} in Appendix~\ref{sc:ThetaLattice}.) 
}
\end{table}

\subsection{Lattice formulation of the non-supersymmetric heterotic string}

In the so-called bosonic or lattice formulation, the supersymmetric E$_8\times$E$_8$ string can be written as 
\begin{equation} \label{SO(16)_lattice_partition} 
{\mathbf Z}_{\text{E}_8^2}(\tau,\overline\tau) = 
{\mathbf Z}_8^x(\tau,\overline\tau) \cdot 
{\boldsymbol{\Gamma}}_4(\tau) \cdot 
\overline{{{\mathbf \Gamma}}_{16}(\tau)}~, 
\end{equation} 
using the lattice partition functions defined in \eqref{Space-time_partition} and \eqref{E8_partition}. 
The N=0 theory can be obtained by considering a (freely acting) $\Intr_2$ orbifold 
of the supersymmetric theory~\cite{Dixon:1986iz,Dixon:1986jc} with twist $v_0$ and shift $V_0$
conveniently chosen as
\begin{equation}\label{WittenTwist}
v_0 = \big(0, 1^3\big)~,
\qquad 
V_0 = \big(1,0^7\big)\big(\sm1,0^7\big)~. 
\end{equation}
Even though this corresponds to a trivial $2 \pi$ space-time twist on three $\Real^2$ planes simultaneously, it does not act trivially on the fermions of the theory~\cite{Rohm:1983aq}.
The modular invariant partition function associated to this orbifolding reads 
\begin{equation} \label{N=0_partition} 
Z_\text{latt}(\tau,\overline\tau) = 
\frac 12 \sum_{l,l'} \, Z_8^x(\tau,\overline\tau) \cdot 
{\mathbf{\Gamma}}_4\brkt{lv_0}{l'v_0}
(\tau) \cdot 
\overline{\mathbf{\Gamma}_{16}\brkt{l V_0}{l' V_0}(\tau)}~. 
\end{equation}
Since both $v_0$ and $V_0$ are vectors that only contain integers, they can be removed from the powers of $\bar q$ and $q$ by shifting the internal summation vectors $n_4\in\Intr^4$ and $n_8,n_8'\in\Intr^8$ in these lattice partition functions~\eqref{Space-time_partition} and~\eqref{E8_partition} with~\eqref{fermion_partition} over appropriate integral vectors. 
However, this gives rise to additional phases from the factors that implement the $\Intr_2$ orbifold projection. 
Hence, the $\Intr_2$ summation variables only appear in such phase factors. 
The sum over $l'$ implements a projection, which leads to a unique solution for the other summation variable $l$. 
After some algebra one can indeed rewrite this partition function in the same form as in~\eqref{SO(16)_partition} with the torsion phases~\eqref{TorsionPhases}. 
\footnote{Applying an analogous twist to the supersymmetric Spin(32)$/\Intr_2$ theory instead, the same non-supersymmetric heterotic theory can be obtained~\cite{Dixon:1986iz}. }

In Table~\ref{tb:Lattices_E8_N=0} we have indicated the effects of the $\Intr_2$ twist on the lattices which define the supersymmetric E$_8\times$E$_8$ theory: 
The entries with gray background in the second column of this Table indicate the lattices which are projected out by this $\Intr_2$ twist. 
The remaining entries with white background thus define the untwisted sectors in the N=0, SO(16)$\times$SO(16) theory. 
The twisted sector lattices, which replace the projected out lattices of the N=1 theory, are indicated by the gray entries in the last column. 
To summarize, the white / gray background entries in the last column of Table~\ref{tb:Lattices_E8_N=0} correspond to the untwisted / twisted sectors ($l=0,1$) of the supersymmetry breaking twist~\eqref{WittenTwist}.

\section{Orbifold compactifications}
\label{sc:Orbifold}

In the previous section we recalled two equivalent descriptions of the non-supersymmetric heterotic SO(16)$\times$SO(16) string in ten dimensions. 
In this section we investigate orbifold compactifications~\cite{Dixon:1985jw,Dixon:1986jc,ibanez_88,Font:1988tp} of this non-supersymmetric theory. 
(For related work see e.g.\ Refs.~\cite{Taylor:1987uv,Toon:1990ij,Sasada:1995wq,Font:2002pq}.)
To have an exact string description, we will extend either of these 
formulations to include the effects of orbifolding, i.e.\ the orbifold projection and the introduction of twisted states. 
Even though the theory in ten dimensions is non-supersymmetric, we will consider its compactification on Calabi-Yau orbifolds only. 
In the general description we present here we will only focus on $T^6/\Intr_N$ orbifolds; 
extensions to $\Intr_M\times \Intr_N$ are possible and will be considered in the model searches we discuss later in this work. 
We first describe non-compact orbifolds for simplicity.
The extension to toroidal orbifolds with possible Wilson lines that distinguish various fixed points is in principle straightforward though notationally tedious; 
in Subsection~\ref{sc:Extension} we quote the results relevant for spectra computations.

\subsection{Non-supersymmetric orbifolds in the generalized-torsion formalism}

The $\Intr_N$ orbifold action is encoded in the following boundary conditions 
\begin{equation} 
X^i(\sigma+1) = e^{2\pi i k v_i}\, X^i(\sigma)~, 
\quad 
\psi^i(\sigma+1) = e^{2\pi i (\frac s2 + k v_i)} \, \psi^i(\sigma)~, 
\quad 
\lambda^I_a(\sigma+1) = e^{2\pi i (\frac {t_a}2 +k V_a^I)}\, \lambda^I_a(\sigma)~, 
\label{eq:coundaryconditions2}
\end{equation}
with $a=1,2$ ($t_1 =t, t_2=u$ when comparing with~\eqref{eq:coundaryconditions1}), $I=1,\ldots, 8$ and $k=0,\ldots N-1$ 
labels the various orbifold sectors. 
Here $X^i$, $i=0,1,2,3$, are complexified coordinates and $\psi^i$ their supersymmetric partners. 
The index $i=0$ refers to the two light-cone uncompactified Minkowski directions. 
The action of the orbifold in the internal space is defined by $v$, whereas the shift $V=\big(V_1;V_2\big)$ encodes its embedding into the gauge degrees of freedom.

We restrict ourselves to twists that would preserve at least N=1 supersymmetry in compactifications
of the N=1 heterotic strings, i.e.\ we take\footnote{We do not indicate the component of $v$ in the four dimensional Minkowskian directions as it is simply zero.}
\begin{equation} \label{orbifold_shift}
v = \big(v_1,v_2, -v_1-v_2\big)~, 
\qquad 
N\, v_1 \equiv N\, v_2 \equiv 0~, 
\end{equation} 
such that the sum of the entries of $v$ is identically zero. 
($\equiv$ means equal modulo integers.) 
For now we assume that the orbifold twist acts non-trivially on all three complex internal directions simultaneously, i.e.\ 
$v_{1,2}, v_1+v_2 \neq 0$ modulo integers.

\subsubsection*{Partition function}

The full partition function for the orbifolded SO(16)$\times$SO(16) theory can be determined from the representation of the partition function given in~\eqref{SO(16)_partition} by including the appropriate shifts of the characteristics to  incorporate the modified boundary conditions due to the orbifolding. 
A choice for the orbifolded partition function is given by
\begin{equation} \label{SO(16)_orbifold_partition} 
{\mathbf Z}_\text{ferm}(\tau,\overline{\tau}) = -\frac 1{8N} \sum T\cdot {\mathbf Z}_2^x(\tau,\overline{\tau}) \cdot 
{\mathbf Z}_6^X\brkt{kv}{k'v}(\tau,\overline{\tau})\cdot 
\widehat {\mathbf Z}_4\brkt{\frac s2e_4 +k v}{\frac{s'}2e_4+k'v}(\tau) \cdot 
 \prod_{a=1,2}
 \overline{\widehat {\mathbf Z}_8\brkt{\frac{t_a}2e_8 + k V_a}{\frac{t_a'}2e_8+k' V_a}(\tau)}~. 
\end{equation} 
for $(k,k') \neq 0$. 
The partition function in the $(k,k') = 0$ sector is just $1/N$ times~\eqref{SO(16)_partition}.

By construction, this orbifold partition function is modular invariant. 
However, we need to ensure that the partition function respects the periodicities of the various labels, i.e.\ $s\sim s+2$, $t_a \sim t_a +2$ and $k \sim k+ N$ and their primed versions. 
Because the additional torsion phase $T$ of the N=0 theory respects these periodicities, the conditions on the orbifold boundaries encoded in $v$ and $V =(V_1; V_2)$ are the same as for orbifolds of the supersymmetric E$_8\times$E$_8$ theory. 
In detail, the three periodicities of the spin-structures are respected provided that 
\begin{subequations} \label{ModInv_ferm_tot}
\begin{equation} \label{LinModInv_ferm} 
\frac 12\, e_4^T v \equiv 
\frac 12\, e_8^T V_1 \equiv 
\frac 12\, e_8^T V_2 \equiv 0~. 
\end{equation} 
For the periodicity of $k$ one must require that 
\begin{equation} \label{ModInv_ferm}
\frac N4\, e_4^T v \equiv 
\frac N4\, e_8^T V_1 \equiv 
\frac N4\, e_8^T V_2 \equiv 0~, 
\qquad 
\frac N2\, \big( v^2 - V^2 \big) \equiv 0~. 
\end{equation} 
\end{subequations} 

\subsubsection*{Mass spectrum}

Given the partition function for the various sectors, it is straightforward to 
determine the mass spectrum of the theory and in particular the massless states. 
One reads off the right- or left-moving masses $M_{L/R}^2$ by making the $\bar q$ or $q$ expansion of the partition function. 
The right-moving mass is given by
\begin{equation}\label{right_moving_p}
M_R^2 = 
\frac 12\, p_\text{sh}^2 
- \frac 12 + \delta c_k + \mathbf{N}_R~, 
\qquad 
p_\text{sh} =n + \frac {s-1}2 e_4 + k\, v~,  
\quad 
n \in \mathbbm{Z}^4~. 
\end{equation}
Here $\delta c_k$ is the familiar vacuum shift 
\begin{equation}
\delta c_k = \frac 12 \tilde v_k^T\big(e_4 - \tilde v_k\big)~,  
\end{equation}
with $\tilde v_k \equiv k v$ such that all the entries fulfill $0 \leq (\tilde v_k)_i < 1$. 
Furthermore, the right-moving number operator $\mathbf{N}_R$ encodes possible right-moving oscillator excitations. 
The left-moving mass reads
\begin{equation}\label{left_moving_p}
M_L^2 = 
\frac 12\, P_\text{sh}^2 
- 1 + \delta c + \mathbf{N}_L~, 
\quad 
P_\text{sh} = \big(P_1;P_2\big)~, 
\quad 
P_a =n_a + \frac {t_a-1}2 e_8 + k\, V_a~,
\quad 
n_a \in \mathbbm{Z}^8~,  
\end{equation}
where $\mathbf{N}_L$ denotes possible left-moving oscillators.

Only (massless) states that survive all generalized GSO and orbifold projections are part of the spectrum.
The GSO projections are modified in the N=0 theory: 
\begin{equation}\label{GSO_s}
\frac 12\, e_4^T n \equiv 
\frac 12\big(t_1+t_2-1 \big)~, 
\end{equation}
and 
\begin{equation}\label{GSO_t}
\frac 12\, e_8^T n_1 \equiv 
\frac 12 \Big( s + t_1 + t_2 + \frac k2 e_8^T V_1 -1\Big)~, 
\qquad 
\frac 12\, e_8^T n_2 \equiv
\frac 12 \Big(s + t_1 + t_2 + \frac k2 e_8^T V_2 -1\Big)~. 
\end{equation}
In particular we see that in this description the GSO projections depend on the sectors of the non-symmetric theory one is considering. 
Because of the constraints on the input parameters given in \eqref{LinModInv_ferm} these are indeed $\mathbbm{Z}_2$ projection conditions. 
Finally, the orbifold projection is the conventional one:
\begin{equation}\label{Orbifold_projection}
V^T P_\text{sh} - v^T p_\text{sh} \equiv 
\frac k2 \big( v^2 - V^2\big) 
 -  \frac {t_1}4 e_8^TV_1 -  \frac {t_2}4 e_8^TV_2~. 
\end{equation}

\subsection{Non-supersymmetric orbifolds in the lattice formulation}

Alternatively, one can describe orbifoldings of the N=0 theory in the lattice formulation.

\subsubsection*{Partition function}

In this language the orbifolded version of the partition function~\eqref{N=0_partition} becomes 
\begin{equation} \label{N=0_orbifold_partition} 
{\mathbf Z}_\text{latt}(\tau,\overline\tau) = 
\frac 1{2N} \sum\, 
 {\mathbf Z}_2^x(\tau,\overline{\tau}) \cdot 
{\mathbf Z}_6^X\brkt{kv}{k'v}(\tau,\overline{\tau})  \cdot 
\widehat{\mathbf{\Gamma}}_4\brkt{l v_0 + k v}{l' v_0+kv}(\tau) \cdot 
\overline{\widehat{\mathbf{\Gamma}}_{16}\brkt{l V_0+k V}{l' V_0+k' V}(\tau)} ~,  
\end{equation} 
using the notation defined in \eqref{lattice_partition_orbifolded}, where $l=0,1$ and as usual $k=0,...,N-1$. 
In this formulation the conditions for modular invariance and proper orbifold and GSO projections read 
\begin{equation} \label{ModInvWitten} 
N\, V_a \in \mathbf{E}_8~, 
\quad a =1,2~, 
\qquad 
\frac {N}2 \,\big(V^2-v^2\big) \equiv V_0 \cdot V \equiv 0~. 
\end{equation} 
These conditions are different and generically weaker than those we obtained in the other description. 
This can be understood as follows: In the lattice formulation the projections that define the space-time lattice $\mathbf{\Gamma}_4$ and gauge lattice $\mathbf{\Gamma}_{16}$ are not modified. 
Hence, one does not have to impose additional constraints to ensure that these projections are well-defined. 
Even though these two descriptions are not equivalent, they are related to each other by additional generalized discrete torsion phase
\begin{equation} 
T_{\text{ferm}\leftrightarrow \text{latt}} = e^{2\pi i\, \frac 14 e_4^Tv (k s' - k' s)}\, 
e^{-2\pi i\, \frac 14 e_8^TV_1 (k t' - k' t)}\, 
e^{-2\pi i\, \frac 14 e_8^TV_2(k u' - k' u)}\, 
e^{2\pi i\, k e_{16}^TV}~. 
\end{equation}
Clearly, these phases only define proper $\Intr_2$ GSO and $\Intr_N$ orbifold 
projections and the last factor drops out when the conditions \eqref{ModInv_ferm_tot} are fulfilled.

Since the conditions in the lattice formulation are weaker, we will primarily use the lattice formulation for our model searches. 
In the cases where both formulations are equivalent, the fermionic formulation provides important consistency checks.

\subsubsection*{Mass spectrum}

An additional advantage of the lattice formulation is that the equations 
that define the massless spectra are essentially the same as those in orbifolds of the supersymmetric theory. 
Concretely, the right- and left-moving mass are given by
\begin{equation}
\label{eqs:mass-equations}
M_R^2 = 
\frac 12\, p_\text{sh}^2 
- \frac 12 + \delta c + \mathbf{N}_R~, 
\qquad 
M_L^2 = 
\frac 12\, P_\text{sh}^2   
- 1 + \delta c + \mathbf{N}_L~, 
\end{equation}
where 
\begin{subequations} \label{ShiftedMomenta} 
\begin{align}
p_\text{sh} &=n + \frac {s-1}2 e_4 + v_g~, 
& v_g &=  l\, v_0+ k\, v~,  
\\[1ex]  \label{GaugeShiftedMomentum} 
P_\text{sh}&=\big(P_1;P_2\big)~, \quad 
P_a =n_a + \frac {t_a-1}2 e_8 + V_{ga}~, 
&
V_g &=  l\, V_{0} + k\, V~, 
\end{align}
\end{subequations}
where the label $g=(l,k)$ refers to the $(l,k)$-sector of the string state. 
The integral vectors $n \in \mathbbm{Z}^4$ and $n_a \in \mathbbm{Z}^8$ are constrained by the standard GSO projections: $\frac 12\, n^Te_4\equiv \frac s2$, $\frac 12\, n_a^Te_8\equiv0$. 
This spectrum is subject to the projection conditions:
\begin{equation} \label{eqs:orbifold_projections_noncompact}
v_0^T  p_\text{sh} - V_0^T P_\text{sh}  \equiv 
\frac 12\, v_0^T\,v_g - \frac 12\, V_0^T\, V_g~, 
\qquad 
v^T  p_\text{sh} - V^T P_\text{sh}  \equiv 
\frac 12\, v^T\, v_g - \frac 12\, V^T\,v_g~. 
\end{equation}

\subsection{Extension to toroidal orbifolds with Wilson lines}
\label{sc:Extension} 

The mass equations~\eqref{eqs:mass-equations}
and projection conditions~\eqref{eqs:orbifold_projections_noncompact} can be 
readily extended to compact toroidal orbifolds including Wilson lines: 
As described above in detail the non-supersymmetric theory can be thought of as a specific $\Intr_2$ orbifold and the extension of non-compact to compact orbifolds is well-known, see e.g.\ \cite{Dixon:1986jc}. 
Therefore, we only quote the crucial modifications here. 

In the presence of Wilson lines $A^\alpha$, the states 
localized at the different fixed points, labeled by the integers $m_\alpha$, are still characterized by solutions to the mass equations~\eqref{eqs:mass-equations} but with the $V_g$ in the shifted momenta $P_\text{sh}$ in~\eqref{GaugeShiftedMomentum} modified to 
\(
V_g =  l\, V_{0} + k\, V + m_\alpha\, A^\alpha,
\)
where the label $g=(l,k,m_\alpha)$ now also indicates at which fixed point (labeled by $m_\alpha$) the state is localized. 
The orbifold projection conditions for each constructed state can be compactly stated as 
\begin{equation}
v_{g'}^T\, p_\text{sh} - V_{g'}^T \,P_\text{sh}
\equiv 
\frac 12\, v_{g'}^T\,v_g - \frac 12\, V_{g'}^T\,V_g~, 
\label{eqs:orbifold_projections_compact}
\end{equation} 
where the label $g'=(l',k',m'_\alpha)$ adopts some adequate integer values for projection conditions (for a detailed explanation, see e.g.~\cite{Ploger:2007iq}). 
In addition, we amend~\eqref{ModInvWitten} with the standard modular invariance conditions between the various Wilson lines, as well as
\begin{equation}
V_0 \cdot A_\alpha \equiv 0~,
\qquad
\alpha=1,\ldots,6~.
\end{equation}
We will use these expressions in the model searches discussed in Section~\ref{sc:Models}.

\subsubsection*{Example: Tachyon--free $\boldsymbol{T^6/\Intr_3}$ orbifold models}

\begin{table}
\begin{center}
\renewcommand{\arraystretch}{1.2}
\scalebox{.9}{
\begin{tabular}{|c||c|}
\hline 
Orbifold shift $V$  & Massless spectrum on orbifold: 
\\ 
Gauge group $G$ & chiral fermions /  \colorbox{lightgray}{ complex bosons } 
\\ \hline\hline 
$\frac 13\big(0,1^2,\sm 2,0^4\big)\big(0^8\big)$  & 
$3 ( \mathbf{3}, \mathbf{1}; \mathbf{16}) 
+ 3 (\overline{\mathbf{3}}, \overline{\mathbf{16}};  \mathbf{1}) 
+  27 (\mathbf{1}, \overline{\mathbf{16}};  \mathbf{1})
+ (\mathbf{1}, \overline{\mathbf{16}} ;  \mathbf{1})   
+ (\mathbf{1}, \mathbf{16};  \mathbf{1})$  
\\ 
& 
$+ (\mathbf{1}; \mathbf{128})
+ (\mathbf{1}, \mathbf{10}; \mathbf{16})
+ 27 (\mathbf{1}; \mathbf{16})$
\\ 
U(3)$\times$SO(10)$\times$SO(16)' & \cellcolor{lightgray}
$81 (\overline{\mathbf{3}},\mathbf{1}; \mathbf{1})
+ 3 (\mathbf{3},\mathbf{1}; \mathbf{1})
+ 3 ( \mathbf{3},\mathbf{10}; \mathbf{1})
+ 27 (\mathbf{1}; \mathbf{1})
+ 27 (\mathbf{1},\mathbf{10}; \mathbf{1})$
\\ \hline\hline 
$\frac 13\big(1^6,0^2\big)\big(1^6,0^2\big)$ & 
$3 (\overline{\mathbf{6}},\mathbf{2}_-; \mathbf{1})
+ 3 (\mathbf{1}; \overline{\mathbf{6}},\mathbf{2}_-)
+ 3 (\mathbf{15},\mathbf{2}_+; \mathbf{1})
+ 3 (\mathbf{1}; \mathbf{15},\mathbf{2}_+) 
+ 3 (\overline{\mathbf{6}},\mathbf{1}; \overline{\mathbf{6}},\mathbf{1})+ 
$
\\
 & 
$3 (\mathbf{1},\mathbf{4}; \mathbf{6},\mathbf{1}) 
+ 3 (\mathbf{6},\mathbf{1}; \mathbf{1},\mathbf{4}) 
+  (\mathbf{20},\mathbf{2}_-; \mathbf{1}) 
+  (\mathbf{1}; \mathbf{20},\mathbf{2}_-) 
+ (\mathbf{1},\mathbf{4}; \mathbf{1},\mathbf{4})+$
\\ & 
$29 (\mathbf{1}; \mathbf{1},\mathbf{2}_+) 
+ 29 (\mathbf{1},\mathbf{2}_+; \mathbf{1})  
+ (\mathbf{6},\mathbf{1}; \overline{\mathbf{6}},\mathbf{1})
+ (\overline{\mathbf{6}},\mathbf{1}; \mathbf{6},\mathbf{1})
+ 27 (\mathbf{1},\mathbf{2}_-; \mathbf{1},\mathbf{2}_-) $
\\ 
U(6)$\times$SO(4)$\times$U(6)'$\times$SO(4)' & \cellcolor{lightgray}
$ 3 (\overline{\mathbf{15}},\mathbf{1}; \mathbf{1}) 
+ 3 (\mathbf{1}; \overline{\mathbf{15}},\mathbf{1})
+ 3 (\mathbf{6},\mathbf{4}; \mathbf{1})
+ 3 (\mathbf{1}; \mathbf{6},\mathbf{4})
+ 27 (\mathbf{1},\mathbf{2}_+;\mathbf{1},\mathbf{2}_+) 
+ 27 (\mathbf{1}; \mathbf{1})$
\\ \hline\hline 
$\frac 13 \big(1^8\big)\big(1^4,0^4\big)$ 
& $3 (\mathbf{8}; \mathbf{1},\mathbf{8}_c)
+ 3 (\mathbf{1}; \mathbf{1},\mathbf{8}_s)
+ 3 (\mathbf{1}; \mathbf{4},\mathbf{8}_v)
+ 3 (\overline{\mathbf{28}};\mathbf{1})
+ 3 (\overline{\mathbf{8}}; \overline{\mathbf{4}},\mathbf{1})
 + (\mathbf{70}; \mathbf{1})$
\\ 
 & 
$+ (\mathbf{1}; \mathbf{6},\mathbf{8}_s)
+ 27 (\mathbf{1}; \mathbf{1},\mathbf{6})
+ 81 (\mathbf{1}; \mathbf{1})
+ 3 (\mathbf{1}; \mathbf{1}) 
+ (\mathbf{8}; \overline{\mathbf{4}},\mathbf{1})
+ (\overline{\mathbf{8}}; \mathbf{4},\mathbf{1})$
\\ 
U(8)$\times$U(4)'$\times$SO(8)' & \cellcolor{lightgray} 
$
 3 (\overline{\mathbf{28}};\mathbf{1})
+ 3 (\mathbf{1}; \mathbf{6},\mathbf{1})
+ 3 (\mathbf{1}; \mathbf{4},\mathbf{8}_c)
+ 27 (\mathbf{1}; \mathbf{1},\mathbf{8}_s)
+ 27 (\mathbf{1};\mathbf{1})
 $
\\ \hline 
\end{tabular}}
\renewcommand{\arraystretch}{1}
\end{center}
\caption{\label{tb:Z3Orbifold_Spectra} 
Massless spectra of some $T^6/\Intr_3$ sample orbifold models with N=0. 
We use the notation $\mathbf{2}_{+/-}$ to indicate the fundamental representations of the two 
SU(2) factors in SO(4) $\cong$ SU(2)$_+\times$SU(2)$_-$. Further, $(\mathbf1;\mathbf1)$ denotes
a singlet under all non-Abelian gauge factors.
}
\end{table}

We consider simple models from $T^6/\Intr_3$ Calabi-Yau orbifold of the N=0 theory to illustrate the general features of orbifolds of the non-supersymmetric theory. 
The $\Intr_3$ orbifold has twist $v= \frac 13(1,1,-2)$, leading to the modular invariance condition
\begin{equation}
\frac 32\, V^2 \equiv \frac 32\, v^2 = 1~. 
\end{equation} 
In Table~\ref{tb:Z3Orbifold_Spectra} we note some sample models resulting from solutions to this equation. 
In this table we give the emerging gauge group and the bosonic and fermionic spectrum. 
The fermionic component of these spectra are free of non-Abelian anomalies and there 
is always a single anomalous U(1) which is universal, i.e. it satisfies~\cite{Casas:1987us}
\begin{equation} \label{universalU(1)}
\frac 1{24}\, \tr\, Q_\text{anom} = \frac 1{6 t_\text{anom}^2} \, \tr\, Q_\text{anom}^3 
= \frac 1{2 t_j^2}\,  \tr \Big(Q_\text{anom} Q_j^2 \Big)=  \tr \Big( \ell(G_i)\, Q_\text{anom} \Big)~, 
\end{equation}
where $Q_\text{anom} = t_\text{anom}^I\, H_I$ is the anomalous U(1), 
$Q_j = t_j^I\, H_I$ possible additional U(1) generators, and $\ell(G_i)$ 
is the Dynkin index of the corresponding representations w.r.t.\ the 
non-Abelian gauge group factor $G_i$.\footnote{Our 
conventions are such that $\ell=1/2$ for the fundamental representation of $G_i=\,$SU($N$).}
We use this equation as a consistency check for all the non-supersymmetric 
four dimensional models we generate in this work.

\subsection{Tachyons in twisted sectors}
\label{sc:orbifold_tachyon_model}

\begin{table}[]
\begin{center}
\renewcommand{\arraystretch}{1.3}
\scalebox{1}{
\begin{tabular}{|l|c|c||l|c|c|}
\hline 
\multicolumn{1}{|c|}{Orbifold} & Twist & Tachyons & \multicolumn{1}{|c|}{Orbifold} & Twist & Tachyons 
\\ \hline\hline 
$T^6/\Intr_3$ & $\frac13(1,1,-2)$ & forbidden &
$T^6/\Intr_2\times\Intr_2$ & $\frac12(1,-1,0)\,;~ \frac12(0,1,-1)$ & forbidden
\\ \hline 
$T^6/\Intr_4$ & $\frac14(1,1,-2)$ & forbidden &
$T^6/\Intr_2\times\Intr_4$ & $\frac12(1,-1,0)\,;~ \frac14(0,1,-1)$ & possible
\\ \hline 
$T^6/\Intr_\text{6-I}$ & $\frac16(1,1,-2)$ & possible &
$T^6/\Intr_2\times\Intr_\text{6-I}$ & $\frac12(1,-1,0)\,;~ \frac16(1,1,-2)$ & possible
\\ \hline 
$T^6/\Intr_\text{6-II}$ & $\frac16(1,2,-3)$ & possible & 
$T^6/\Intr_2\times\Intr_\text{6-II}$ & $\frac12(1,-1,0)\,;~ \frac16(0,1,-1)$ & \cellcolor{lightgray} possible
\\ \hline 
$T^6/\Intr_7$ & $\frac17(1,2,-3)$ & possible &
$T^6/\Intr_3\times\Intr_3$ & $\frac13(1,-1,0)\,;~ \frac13(0,1,-1)$ & possible 
\\ \hline 
$T^6/\Intr_\text{8-I}$ & $\frac18(1,2,-3)$ & possible &
$T^6/\Intr_3\times\Intr_6$ & $\frac13(1,-1,0)\,;~ \frac16(0,1,-1)$ & \cellcolor{lightgray} possible
\\ \hline 
$T^6/\Intr_\text{8-II}$ & $\frac18(1,3,-4)$ & possible & 
$T^6/\Intr_4\times\Intr_4$ & $\frac14(1,-1,0)\,;~ \frac14(0,1,-1)$ & possible
\\ \hline 
$T^6/\Intr_\text{12-I}$ & $\frac1{12}(1,4,-5)$ & possible & 
$T^6/\Intr_6\times\Intr_6$ & $\frac16(1,-1,0)\,;~ \frac16(0,1,-1)$ & \cellcolor{lightgray} possible 
\\ \hline 
$T^6/\Intr_\text{12-II}$ & $\frac1{12}(1,5,-6)$ & \cellcolor{lightgray} possible  
\\ \hhline{---}
\end{tabular}}
\renewcommand{\arraystretch}{1}
\end{center}
\caption{\label{tb:TachyonicOrbifolds}
This table indicates for which Abelian Calabi-Yau orbifolds twisted ground state tachyons are possible or strictly forbidden when used for compactification of the non-supersymmetric heterotic SO(16)$\times$SO(16) theory. The grey background entries flag orbifolds in which additional excited right-moving tachyons might appear in some sectors.}
\end{table}

In the ten dimensional theories the right-mover could become tachyonic if its weight comes from the root lattice. 
In the N=0 theory this happens in the $\mathbb{Z}_2$ twisted sectors, but the left-mover side does not allow to level-match this tachyon as its underlying lattice is $\mathbf{V}_8 \otimes \mathbf{C}_8 \oplus \mathbf{C}_8 \otimes \mathbf{V}_8$ which leads only to massive states. 
However, when we compactify on toroidal orbifolds, this is no longer the case, i.e.\ there can be shift vectors which shift the lattice of the left-movers such that tachyonic level-matching can be achieved. 

Let us develop some criterion to determine from which sector tachyonic levels for the right movers could arise in a given geometry.
Since the untwisted sectors are obtained by projections on the ten-dimensional spectrum, they are guaranteed to be free of level-matched tachyons. 
Hence, tachyons can only arise in twisted sectors. 
We consider a twisted sector $\mathbb{N} \ni k \neq 0$ where the twist has the form provided in \eqref{orbifold_shift} except that now we allow for zero entries.  
In the sectors where $p_\text{sh}$ comes from the shifted vector, spinor or cospinor lattice, the mass equation \eqref{right_moving_p} implies that the lightest states are precisely massless. 
However, if the shifted momentum comes from the 10D root lattice, tachyonic right movers 
may appear. 
The properties of these tachyons is determined by the shortest weight $p_\text{sh}$ 
solving~\eqref{right_moving_p} for some negative value of $M_R^2$. 
This $p_\text{sh}$ can be identified by first noting that we can always rearrange the components so that $p_\text{sh}=(\omega_1, \omega_2, -\omega_1-\omega_2)$ with 
$0 \leq \omega_1 \leq \omega_2 \leq \frac12$, by adding some root lattice 
vector to $k v$. 
Then, substituting this into \eqref{right_moving_p}, we find
\begin{equation}\label{TachyonCriterium}
M_R^2 
= \omega_1 + \omega_2 - \frac12 \hspace{0.5cm}\text{with }~ \omega =  k v + q~, \hspace{0.2cm} q \in \text{SO(8)}_R
\end{equation}
in the absence of right-moving oscillators, i.e.\ $\mathbf{N}_R=0$.
In each sector there is at most one $p_\text{sh}$ to satisfy \eqref{TachyonCriterium} with negative $M_R^2$.
The CPT conjugates of these tachyonic states come from the $N-k$ sectors as usual with $p_\text{sh}\rightarrow -p_\text{sh}$.

In Table~\ref{tb:TachyonicOrbifolds} we indicate which orbifolds with supersymmetric geometries have twisted right-moving tachyons. 
In these cases also states with right-moving oscillators contribute to the massless spectrum unlike in compactifications of the N=1 theory. 
The orbifolds $T^6/\Intr_\text{12-II}$, $T^6/\Intr_2\times\Intr_\text{6-II}$, $T^6/\Intr_3\times\Intr_6$ and $T^6/\Intr_6\times\Intr_6$ even have tachyonic states with right-moving oscillators.

\subsubsection*{Example: Tachyonic $\boldsymbol{T^6/\Intr_\text{6-I}}$ orbifold}

We consider a $\Intr_\text{6-I}$ orbifold on the factorized SU$(3)^3$ lattice of the N=0 theory, with shift vector
\begin{equation} \label{TachyonicShiftExample}
%
%
V = \frac 16
\big( -2, -16,  -14,  -2,    2,     6,   3,  11\big) \big( -2,  -5,    -6,  -2,      6, -13,  -1,  19\big)~
%
%
\end{equation}
and no Wilson lines. 
This leads to a non-Abelian gauge group SU(5)$\times$SU(4)'$\times$SO(4)'$\times$SU(2)' together with six U(1) factors. 
The full spectrum is shown in Table~\ref{tab:TachyonicOrbifold}. 
The shift allows to level-match the right moving tachyon from the $(s,t,u) = (1,0,1)$ sector, cf.~Table~\ref{tb:Lattices_E8_N=0}. 
More precisely, this tachyon comes with multiplicity three from the three $\Intr_\text{6-I}$ fixed points, and transforms as a $(\mathbf{1};\mathbf{1},\mathbf{1},\mathbf{2})$. 
This tachyon potentially leads to an instability of the theory which would drive it away from the orbifold point and at the same time it would break parts of the gauge group. 
In Section \ref{sec:blowup_tachyon_model} we will discuss how the blow-up of this model removes the tachyon, showing that to lowest order in perturbation theory and $\alpha'$-expansion this model could have a stable vacuum.

\begin{table}[]
\begin{center}
\renewcommand{\arraystretch}{1.2}
\scalebox{.95}{
 \begin{tabular}{|c||c|}
 \hline
 States &  Gauge representations of the spectrum of a tachyonic $\Intr_\text{6-I}$ orbifold \\
 \hline \hline 
\cellcolor{lightgray} Bosonic tachyons & \cellcolor{lightgray}
  $3 (  \mathbf{1};   \mathbf{1},   \mathbf{1},   \mathbf{2})$
 \\ \hline\hline  
 Massless & 
 $ 4 (  \mathbf{10};   \mathbf{1}) +  (  \overline{\mathbf{10}};   \mathbf{1}) 
 + 6 ( \mathbf{5};   \mathbf{1})  + 3 ( \overline{\mathbf{5}};   \mathbf{1}) 
 + (   \mathbf{5};   \mathbf{1},   \mathbf{4},   \mathbf{1})
 + 2 ( \overline{ \mathbf{5}};   \mathbf{1},   \mathbf{1},   \mathbf{2})
 + ( \mathbf{5};   \mathbf{1},   \mathbf{1},   \mathbf{2})
 $
 \\
 chiral fermions & 
 $+ 2( \overline{ \mathbf{5}};   \mathbf{4},   \mathbf{1},   \mathbf{1})
 + 12 (\mathbf{1};   \mathbf{4},   \mathbf{1},   \mathbf{1})
 + 18  (\mathbf{1};   \overline{\mathbf{4}},   \mathbf{1},   \mathbf{1})
 +  2 (   \mathbf{1}; \overline{ \mathbf{4}},   \mathbf{2}_-,   \mathbf{2}) 
 + 2 (   \mathbf{1};   \mathbf{4},   \mathbf{2}_+,   \mathbf{1})
 $ 
\\ &
$+ ( \mathbf{1};   \mathbf{6},   \mathbf{2}_-,   \mathbf{1})
+ ( \mathbf{1};   \mathbf{6},   \mathbf{2}_+,   \mathbf{1})
+ 12 (\mathbf{1};   \mathbf{1},   \mathbf{2}_+,   \mathbf{2})
+ 4 (\mathbf{1};   \mathbf{1},   \mathbf{4},   \mathbf{1})
+ 36 ( \mathbf{1};   \mathbf{1},   \mathbf{2}_-,   \mathbf{1})$
\\ & 
$ + 30 (\mathbf{1};   \mathbf{1},   \mathbf{2}_+,   \mathbf{1})
+ 11 (\mathbf{1};   \mathbf{1},   \mathbf{1},   \mathbf{2})
+ 53 (   \mathbf{1};  \mathbf{1}) 
$
 \\ \hline\hline  
 \cellcolor{lightgray} Massless & \cellcolor{lightgray} 
$
9 ( \mathbf{5};   \mathbf{1})
+ 2 ( \overline{ \mathbf{5}};   \mathbf{1})
+ (\overline{ \mathbf{10}};   \mathbf{1})
+ ( \mathbf{1};   \mathbf{1},   \mathbf{4},   \mathbf{2})
+ 30 (   \mathbf{1};   \mathbf{1},   \mathbf{2}_-,   \mathbf{1})
+ 12 (  \mathbf{1};   \mathbf{6},   \mathbf{1},   \mathbf{1})$ 
 \\ 
\cellcolor{lightgray} complex scalars & \cellcolor{lightgray}
$+  2 (   \mathbf{1};   \mathbf{4},   \mathbf{1},   \mathbf{2})
+ 2 (   \mathbf{1}, \overline{ \mathbf{4}},   \mathbf{4},   \mathbf{1})
+ 22 (   \mathbf{1};   \mathbf{1},  \mathbf{2}_+,   \mathbf{1}) 
+ 10 (  \mathbf{1};  \mathbf{1},   \mathbf{2}_-,  \mathbf{2}) 
+ 46 ( \mathbf{1};  \mathbf{1})
$
 \\ \hline
 \end{tabular}}
\renewcommand{\arraystretch}{1}
\end{center}
 \caption{Spectrum of the $\Intr_\text{6-I}$ orbifold model defined by shift~\eqref{TachyonicShiftExample}.
 This spectrum contains a tachyon in a non-trivial representation of the gauge group. 
}
\label{tab:TachyonicOrbifold}
\end{table}

\section{Smooth compactifications}
\label{sc:Smooth}

In the previous section we have considered compactifications of the N=0, SO(16)$\times$SO(16) theory on orbifolds which themselves would preserve N=1 supersymmetry when used to compactify N=1 string theories. 
Motivated by these results, we now consider compactification of the low-energy N=0 theory given in Table~\ref{tb:N=0Spectrum} on a smooth Calabi-Yau manifold $\mathcal{M}^6$. 
Like for Calabi-Yau compactification of heterotic supergravity, we assume that we can at least topologically characterize the compactification manifold by its curvature two-form class $\mathcal{R}_2$ and a vector bundle corresponding to a two-form field strength $\mathcal{F}_2$, which satisfy the integrated Bianchi identities 
\begin{equation} 
\int_{\mathcal{C}^4} \left\{ \tr\, \mathcal{R}_2^2 - \tr\, \mathcal{F}_2^2 \right\} = 0~, 
\end{equation} 
for any closed four-cycle $\mathcal{C}^4 \subset \mathcal{M}^6$  (see e.g.\ \cite{Candelas:1985en,gsw_2}). 

\subsubsection*{4D fermionic spectrum}

To compute the four dimensional chiral fermionic spectrum on this background we can rely on conventional techniques to determine the zero modes of the Dirac operator. 
In particular, for line bundle backgrounds we may employ the multiplicity operator~\cite{Nibbelink:2007rd,Nibbelink:2007pn}
\begin{equation}  
\mathcal{N} = \int \left\{
\frac 16\, \Big(\frac{\mathcal{F}_2}{2\pi}\Big)^3 - \frac 1{24}\, \frac{\mathcal{F}_2}{2\pi}\, \tr\, \Big(\frac{\mathcal{R}_2}{2\pi}\Big)^2 \right\}~, 
\end{equation}
which can be thought of as a representation-dependent index. 
For non-Abelian embeddings, one has to take the trace over the resulting representations of the internal group in which $\mathcal{F}_2$ takes its values. 
More elaborate techniques using cohomology theory (see e.g.\ 
\cite{Candelas:1985en,Donagi:2004ub,Bouchard:2005ag,Braun:2005ux,Anderson:2007nc,Anderson:2008uw})
can also be applied. 
Compared to the conventional computation of spectra in heterotic supergravity, there are two novel issues one should keep in mind in the case of the N=0 theory: 
\begin{enumerate}[i.)]  
\item The charged chiral fermions, $\Psi_+$ and $\Psi_-$ come in both ten dimensional chiralities. 
\item These charged fermions do not come from the adjoint representation of the gauge group in ten dimensions, but rather lie in the $(\mathbf{128},\mathbf{1})+(\mathbf{1},\mathbf{128})$ and $(\mathbf{16},\mathbf{16})$ for $\Psi_+$ and $\Psi_-$, respectively. 
\end{enumerate}

\subsubsection*{4D bosonic spectrum}

In compactifications that preserve at least N=1 supersymmetry in four dimensions, we do not need to do any work to determine the massless scalars, because they are always paired up with the chiral fermions in chiral multiplets. 
For generic non-supersymmetric theories, it is much more difficult to get access to the number and properties of the massless scalars as one has to determine the zero modes of the corresponding Laplace operators. 
However, in our case we are considering the compactification of a non-supersymmetric theory on a Calabi-Yau manifold, which by itself does preserve N=1 supersymmetry. 
This we can exploit to determine the spectrum of massless scalars for the Calabi-Yau compactification of the N=0 theory. 

Because the background we consider can preserve N=1 supersymmetry, the multiplicities and representations of zero modes of the bosonic fields in ten dimensions are the same as those one would obtain if one would consider their -- non-existing -- superpartners on this background. 
For these fermionic superpartners, which are the projected out SO(16)$\times$SO(16) gauginos (and gravitino and dilatino), one can apply the known techniques to compute their Dirac operator, like the representation-dependent index mentioned above. 

More generally, at the lowest order in $\alpha'$ and $g_s$ expansion which we are only considering here, the complete structure of the theory of the bosons in four dimensions is as if they were part of a supersymmetric theory. 
In particular, couplings are restricted and the scalar potential is determined by hypothetical D- and F-terms which arise from a hypothetical superpotential $W$. 
Hence, we expect that a lot of phenomenological aspects of such compactifications beyond the zero mode spectra can be exploited by known techniques of Calabi-Yau compactification.

Moreover, based on this we can argue that we will never encounter tachyons on smooth Calabi-Yau compactifications of the N=0 theory when all curves, divisors and the manifold as a whole are large. 
The Laplacian for gauge fields on a Calabi-Yau background is equal to the Dirac operator of the corresponding gauginos squared~\cite{gsw_2}. 
Consequently, the eigenvalues of this Laplacian are non-negative and no tachyonic states are possible. 
However, tachyons could arise for generic non-supersymmetric or singular backgrounds. 
In particular, tachyonic masses may be generated by non-perturbative effects in generic Calabi-Yau compactifications when volumes of certain cycles become small. 
Hence, this suggests that all the tachyons encountered in certain orbifold theories should be understood as such artifects of the blow-down limit. 
To illustrate and test these techniques we consider the N=0 theory on the resolution of $T^6/\Intr_3$ orbifold, Calabi-Yau standard embeddings and the resolution of a $T^6/\Intr_\text{6-I}$ orbifold model that possesses tachyons as examples in the subsections below.

\begin{table}
\begin{center}
\renewcommand{\arraystretch}{1.2}
\scalebox{.9}{
\begin{tabular}{|c||c|}
\hline 
Line bundle vector $W$  & Massless spectrum in blow-up: 
\\ 
Gauge group $G$ & chiral fermions /  \colorbox{lightgray}{ complex bosons }
\\ \hline\hline 
$\frac13 \big(0,2^3,0^4\big)\big(0^8\big)$ & 
$ 3 (\mathbf{3},\mathbf{1}; \mathbf{16})_{2}
+ 3 (\overline{\mathbf{3}},\overline{\mathbf{16}}; \mathbf{1})_1
+ 27 (\mathbf{1},\overline{\mathbf{16}}; \mathbf{1})_{\sm 3}$
\\ 
U(3)$\times$SO(10)$\times$SO(16)' & \cellcolor{lightgray}
$78 (\overline{\mathbf{3}},\mathbf{1};\mathbf{1})_{4} 
+ 3 (\mathbf{3},\mathbf{10};\mathbf{1})_{2}$
\\ \hline\hline 
$\frac13 \big(1^6,0^2\big)\big(1^6,0^2\big)$  & 
$ 3 (\overline{\mathbf{6}},\mathbf{2}_-; \mathbf{1})_{\sm 2} 
+ 3 (\mathbf{1}; \overline{\mathbf{6}},\mathbf{2}_-)_{\sm 2} 
+ 3 (\mathbf{15},\mathbf{2}_+; \mathbf{1})_1 
+ 3 (\mathbf{1}; \mathbf{15},\mathbf{2}_+)_1 
+ 3 (\overline{\mathbf{6}},\mathbf{1}; \overline{\mathbf{6}},\mathbf{1})_{2}$
\\
& 
$ + 3 (\mathbf{6},\mathbf{1}; \mathbf{1},\mathbf{4})_{\sm1} 
+ 3 (\mathbf{1},\mathbf{4}; \mathbf{6},\mathbf{1})_{\sm1} 
+ 27 (\mathbf{1}, \mathbf{2}_+; \mathbf{1})_{\sm 3} 
+ 27 (\mathbf{1}; \mathbf{1},\mathbf{2}_+)_{\sm 3} $
\\ 
U(6)$\times$SO(4)$\times$U(6)'$\times$SO(4)' & \cellcolor{lightgray}
$ 3 (\overline{\mathbf{15}},\mathbf{1}; \mathbf{1})_{2} 
+ 3 (\mathbf{1}; \overline{\mathbf{15}},\mathbf{1})_{2}
+ 3 (\mathbf{6},\mathbf{4}; \mathbf{1})_{\sm1}
+ 3 (\mathbf{1}; \mathbf{6},\mathbf{4})_{\sm1}$
\\ \hline\hline 
$\frac13 \big(1^8\big)\big(1^4,0^4\big)$ 
& $3 (\mathbf{8}; \mathbf{1},\mathbf{8}_v)_{\sm1} 
+ 3 (\mathbf{1}; \mathbf{1},\mathbf{8}_s)_{\sm 2} 
+ 3 (\mathbf{1}; \mathbf{4},\mathbf{8}_c)_{1} 
+ 3 (\overline{\mathbf{28}};\mathbf{1})_{\sm 2} $
\\ 
 & 
$ + 3 (\overline{\mathbf{8}}; \overline{\mathbf{4}},\mathbf{1})_{2} 
+ 78 (\mathbf{1}; \mathbf{1})_{\sm 4}$
\\ 
U(8)$\times$U(4)'$\times$SO(8)' & \cellcolor{lightgray} 
$ 3 (\overline{\mathbf{28}}; \mathbf{1})_{2}
+ 3 (\mathbf{1}; \mathbf{6},\mathbf{1})_{2}
+ 3 (\mathbf{1}; \mathbf{4}, \mathbf{8}_v)_{\sm1}$
\\ \hline 
\end{tabular}} 
\renewcommand{\arraystretch}{1}
\end{center}
\caption{\label{tb:Z3Blowup_Spectra}
Samples of line bundle models on the resolution of the $T^6/\Intr_3$ orbifold. The complete massless spectra of both the chiral fermions and complex bosons are given. 
The U(1) charge indicated here is identified by the line bundle vector $W_{\alpha\beta\gamma}=W$ and relevant for the multiplicities determined by the multiplicity operator in eq.~\eqref{Z3multiplicity}.
}
\end{table}

\subsection{Line bundle models on the resolution of $\boldsymbol{T^6/\Intr_3}$}

The resolution of the $T^6/\Intr_3$ orbifold has been discussed in various works
\cite{Nibbelink:2007rd,Nibbelink:2007xq,Nibbelink:2008tv}, hence we only quote the necessary results briefly here. The resolution is characterized by the following irreducible set of divisors~\cite{Lust:2006zh,Reffert:2007im,Blaszczyk:2011hs}: 
inherited divisors $R_1, R_2, R_3$ corresponding to the three two-tori which define $T^6$, and 
exceptional divisors $E_{\alpha\beta\gamma}$ with $\alpha,\beta,\gamma =1,2,3$ labeling the 27 fixed points of this orbifold. 
Their non-vanish intersection ring is given by  
\begin{equation}
E_{\alpha\beta\gamma}^3 =R_1R_2R_3 = 9~. 
\end{equation}
A generic line bundle background can be expanded as 
\begin{equation} 
\frac {\mathcal{F}_2}{2\pi} =  \sum W^I_{\alpha\beta\gamma}\, E_{\alpha\beta\gamma}\, H_I, 
\end{equation}
where the line bundle vectors which characterize the embedding on the Cartan of the SO(16)$\times$SO(16) gauge group (generated by $H_I$), must be quantized as $1/3$ times only integers or half-integers such that the integrated Bianchi identities are satisfied, i.e.
\begin{equation}
W_{\alpha\beta\gamma}^2 = \frac43~, 
\end{equation} 
for each $\alpha,\beta,\gamma=1,2,3$ separately.  
And the multiplicity operator reads
\begin{equation} \label{Z3multiplicity} 
\mathcal{N} = \frac 12\, \sum_{\alpha,\beta,\gamma}\big(-3 H_{W_{\alpha\beta\gamma}}^2 + 1\big) 
H_{W_{\alpha\beta\gamma}}~, 
\end{equation}
where $H_{W_{\alpha\beta\gamma}}= W^I_{\alpha\beta\gamma} H_I$.

In Table~\ref{tb:Z3Blowup_Spectra} we give the resulting gauge group and fermionic and bosonic spectrum when we take the same line bundle vector for all exceptional cycles, i.e.\ $W_{\alpha\beta\gamma}=W$, for the orbifold models given in Table~\ref{tb:Z3Orbifold_Spectra}. 
It is not difficult to confirm that all irreducible anomalies which the chiral fermions could induce cancel out. 
Like for the resolutions of supersymmetric models, we see that the spectra on the resolution and orbifold can be matched, provided that one takes into account the consequences of the VEVs of the blow-up modes. 
These blow-up modes are easily identified in the orbifold spectrum: They are complete singlets under the non-Abelian part of the gauge group and have multiplicity of 27 due to the 27 fixed points on the $T^6/\Intr_3$ orbifold. 
The chiral fermion states can be paired up by mass terms that involve Yukawa couplings with the blow-up modes. 
For the complex scalars one can in principle always write down mass terms. 
However, we see that precisely the Yukawa couplings, that would show up in the superpotential, give rise to F-term potentials involving the blow-up modes and the states that disappeared from the orbifold spectrum in blow-up.

\subsection{The standard embedding on Calabi--Yaus}

In general Calabi--Yau compactifications of heterotic supergravity, the most prominent technique to obtain four dimensional spectra is to compute the cohomology groups of the vector bundle in the various representations. 
After decomposing the ten dimensional representations into the bundle structure group and its commutant, the dimensions of the cohomology groups are the multiplicities in four dimensions.
In N=0 language, the cohomology dimensions count, on the one hand, the number of complex scalar bosons which result from a ten dimensional vector field and, on the other hand, the number of left-chiral fermions coming from a left-chiral ten dimensional Majorana fermion. 
The same applies also to the right-chiral fermions since the Calabi--Yau only cares about the ``internal chirality''. 

This equal treatment of vectors and left- and right-chiral spinors allows us to apply the same techniques to the charged spectrum in Calabi--Yau compactifications of the SO(16)$\times$SO(16) theory. 
For the standard embedding, the four dimensional gauge group, as commutant of the bundle structure group SU(3), is SO(10)$\times$U(1)$\times$SO(16)$'$. 
The bosonic and fermionic spectra are shown in Table \ref{tab:genCYspectra}  (which can be found in~\cite{Font:2002pq}). 

\begin{table}
\begin{center}
\renewcommand{\arraystretch}{1.2}
\scalebox{1}{
 \begin{tabular}{|c||c|c|}
 \hline
 Multiplicity & \cellcolor{lightgray} Complex bosons & Chiral fermions \\
 \hline \hline 
 1 & \cellcolor{lightgray} $-$ & $ (\mathbf{16};\mathbf{1})_{3}  + (\overline{\mathbf{16}};\mathbf{1})_{\sm3} + (\mathbf{1};\mathbf{128})_0 + (\mathbf{10};\mathbf{16})_0$ \\
 \hline
 $h^{1,1}$ & \cellcolor{lightgray} $(\mathbf{10};\mathbf{1})_2 + (\mathbf{1};\mathbf{1})_{\sm 4} $ & $(\mathbf{16};\mathbf{1})_{\sm 1} + (\mathbf{1};\mathbf{16})_{\sm 2}$ \\
 \hline
 $h^{1,2}$ & \cellcolor{lightgray} $(\mathbf{10};\mathbf{1})_{\sm 2} + (\mathbf{1};\mathbf{1})_{4} $ & $(\overline{\mathbf{16}};\mathbf{1})_{1} + (\mathbf{1};\mathbf{16})_{2}$ \\
 \hline
 $h^{1}(\text{End}(V))$ & \cellcolor{lightgray} $(\mathbf{1};\mathbf{1})_0 $ & $ - $ \\
 \hline
 \end{tabular}}
\renewcommand{\arraystretch}{1}
\end{center}
\caption{Bosonic and fermionic spectra in Calabi--Yau compactification with standard embedding of the N=0 theory. Right chiral fermions in four dimensions are counted as left-chiral fermions with opposite gauge charge. 
\label{tab:genCYspectra}}
\end{table}

Using this spectrum we take a look at the anomalies that can arise. The SO(2$N$) factors are anomaly free, so the only non-vanishing anomalies are of the form U(1)$-G^2$ with $G=$U(1), SO(10), SO(16) and gravity. Furthermore, in Table \ref{tab:genCYspectra} the first universal row is vector-like and as a result all anomalies are proportional to the Euler number of the Calabi--Yau. In addition we find that the coefficients satisfy the anomaly universality condition~\eqref{universalU(1)}.  Thus the total anomaly is always canceled by the universal axion and the U(1) gauge factor becomes massive. It is phenomenologically very appealing that already just using the standard embedding one gets very close to an SO(10) GUT with bosonic 10-plets as potential Higgses and fermionic 16-plets for the standard model matter families.

\subsection{Resolution of a tachyonic $\boldsymbol{T^6/\Intr_\text{6-I}}$ orbifold}
\label{sec:blowup_tachyon_model}

From the compactification of heterotic supergravity on smooth Calabi--Yaus we know that the tree-level scalar potential, which results from the kinetic term of ten dimensional vector bosons is equal to a sum of positive definite terms to lowest order in $\alpha'$.
It furthermore has the property that its minimum is obtained when all fields have zero VEVs. 
Since this applies to the Calabi--Yau compactification of any ten dimensional Yang--Mills theory, we expect that the N=0 theory has no tachyons in this approximation.  
However, in Section \ref{sc:orbifold_tachyon_model}, we gave an example of an orbifold model with a tree-level tachyon. 
The question arises if this can be matched with a blow-up model and what happens to that tachyon.

To set up the resolution model we first give the relevant divisors and their intersections~\cite{Lust:2006zh}. 
There are three exceptional divisors $E_{1,\gamma}$ from the $\mathbb{Z}_6$ sectors, 15 exceptional divisors $E_{2,\mu\gamma}$ from the $\mathbb{Z}_3$ sectors and six exceptional divisors $E_{3,\nu}$ from $\mathbb{Z}_2$ sectors, where $\gamma=1,2,3$, $\mu=2,\ldots,5$ and $\nu=2,\ldots,6$. As usual, there are inherited divisors $R_a$, $a=1,2,3$. 
The non-vanishing intersection numbers are
\begin{equation}
 \begin{array}{c}
R_1 R_2 R_3 = 18~, 
\qquad 
R_3 E_{3,1}^2 = E_{2,1\gamma}E_{3,1}^2 = - E_{1,\gamma}^2E_{2,1\gamma} = -2~, 
\qquad 
R_3 E_{3,\nu}^2 = -6~, 
\\[2ex]
 E_{1,\gamma}^3 =  E_{2,1\gamma}^3 = E_{3,1}^3 = 8~,  
 \qquad 
 E_{2,\mu\gamma}^3 = 9~, 
 \qquad  
 E_{1,\gamma}E_{2,1\gamma}^2 = -4~. 
 \end{array}
\end{equation}
We make the following ansatz for the bundle vectors
\begin{equation}
\frac{\mathcal{F}}{2\pi} = \Big(W_1^I \sum_{\gamma} E_{1,\gamma} + W_2^I \sum_{\gamma\mu} E_{2,\mu\gamma} + W_3^I \sum_{\nu} E_{3,\nu} \Big) H_I \,,
\end{equation}
which treats all fixed points in the same way. 
Then the Bianchi identities simplify to
\begin{equation}
W_1^2 = W_3^2 = \frac32~, 
\qquad  
W_2^2 = \frac43~, 
\qquad  
W_1\cdot W_2 =  W_2\cdot W_3 = \frac13~.  
\end{equation}
The multiplicity operator takes the form
\begin{equation} \label{MultiplicityTachyonicExample}
 \mathcal{N} = 4 H_1^3 + 22 H_2^3 + \frac43 H_3^3 + 3 H_1^2 H_2 - 6 H_1 H_2^2 - 3 H_2 H_3^2 - H_1 - 7 H_2 - \frac13 H_3~,
\end{equation}
with $H_i = W_i^I\,H_I$.

In order to obtain the bundle vectors from the twisted scalar bosons, we choose the bundle vectors
\begin{equation} \label{BundleVectorsTachyonicExample}
\begin{array}{rcl}
%
W_1 &=& \frac16  \big(1, -1, 1, 1,  -1, -3, 0, 2 \big)  \big( 1, -2, -3, 1, 3, 2, 2, -2 \big)~, 
\\[1ex]
W_2 &=& \frac16  \big(  -1,1, -1, -1, 1, 3, -3, 1  \big)  \big( -1, -1, -3, -1, 3, 1, 1, -1 \big)~, 
\\[1ex] 
W_3 &=& \frac12  \big( 0, 0, 0, 0, 0, 0, -1, 1  \big) \big(0, 1, 0, 0,  0, 1, 1, 1 \big)~.
\end{array}
\end{equation}
They correspond to the shifted momenta of twisted scalar fields. The blow-up modes in the $\theta^1$ and $\theta^2$ sectors transform in the $(\rep1;\rep1,\rep2_-,\rep1)$ representations and the $\theta^3$ blow-up mode is a $(\rep1;\rep6,\rep1,\rep1)$. 
This breaks the orbifold gauge group down to SU$(5)\times$SO$(4)'\times$SO$(4)'$. 
Let us stress one important point for later reference: The blow-up mode in the $\theta^2$ sector is uniquely identified.

\begin{table}
\begin{center}
\renewcommand{\arraystretch}{1.2}
\scalebox{.95}{
 \begin{tabular}{|c||c|}
 \hline
 States &  Non-Abelian representations of a blown-up tachyonic orbifold model\\
 \hline \hline 
\cellcolor{lightgray} Bosonic tachyons & \cellcolor{lightgray} none  
 \\ \hline\hline  
 Massless & $3  (\brep{10};\rep{1}) + 3  (\rep{5};\rep{1})+ 6  (\brep{5};\rep{1}) + 2  (\brep{5};\rep{1},\rep{2_+}) + 2  (\rep5;\rep2_-,\rep1) + 2 (\rep5;\rep2_+,\rep1) + (\rep5;\rep1,\rep2_-)$    
 \\
 chiral fermions & $2  (\rep1;\rep4,\rep1)+ 2  (\rep1;\rep1,\rep4)+ 2  (\rep1;\rep2_+,\rep2_+) + 4  (\rep1;\rep2_+,\rep2_-) + 2  (\rep1;\rep2_-,\rep2_+)  $
 \\ 
 & $ 4  (\rep1;\rep2_-,\rep2_-) +  6  (\rep1;\rep2_+,\rep1) + 8  (\rep1;\rep2_-,\rep1) + 34  (\rep1;\rep1,\rep2_+) + 11  (\rep1;\rep1,\rep2_-)+ 53  (\rep1;\rep1)  $ 
 \\
 \hline\hline  
 \cellcolor{lightgray} Massless & \cellcolor{lightgray}  $ (\brep{10};\rep{1}) + 9  (\rep{5};\rep{1})+ 2 (\brep{5};\rep{1}) + 2  (\rep1;\rep4,\rep1)+ 2  (\rep1;\rep1,\rep4)$
 \\ 
\cellcolor{lightgray} complex scalars & \cellcolor{lightgray}  $4  (\rep1;\rep2_+,\rep2_+) + 2  (\rep1;\rep2_+,\rep2_-) + 4  (\rep1;\rep2_-,\rep2_+) + 2  (\rep1;\rep2_-,\rep2_-) + 43 (\rep1;\rep1) $
 \\ \hline
 \end{tabular}}
\renewcommand{\arraystretch}{1}
\end{center}
 \caption{Spectrum of the full resolution of the tachyonic orbifold model given in Table~\ref{tab:TachyonicOrbifold}. 
 The model has been resolved using the line bundles specified in~\eqref{BundleVectorsTachyonicExample}.  
 Note that the tachyonic state of  that table is absent in the full blown-up model.}
\label{tab:TachyonicOrbifoldBlowup}
\end{table}

In Table~\ref{tab:TachyonicOrbifoldBlowup} we give the massless spectrum as determined by the multiplicity operator~\eqref{MultiplicityTachyonicExample}. 
A more detailed analysis, including the U(1) charges, shows that for all massless fields we find perfect agreement with the orbifold spectrum in Table \ref{tab:TachyonicOrbifold}, up to field redefinitions and decoupling of vector-like states as explained in \cite{Groot Nibbelink:2007ew}. 
Whereas in previous supersymmetric compactifications such matchings were performed on the level of chiral superfields, here we extend them to the bosonic and fermionic spectra separately. 
Moreover, the fermions remember if they stem from ten dimensional spinor or cospinor representations although the four dimensional chiralities of spinors can be changed by complex conjugation.

As expected from the general theory developed above, there are no tachyons when the orbifold has been fully resolved to a smooth Calabi-Yau manifold. Therefore, one may wonder what happened to the twisted tachyons of the $T^6/\Intr_\text{6-I}$ orbifold model given in Subsection~\ref{sc:orbifold_tachyon_model}. 
To figure this out, we focus on the following bosonic fields suppressing the fixed point multiplicities: 
\begin{center}
\renewcommand{\arraystretch}{1.2}
 \begin{tabular}{|cl|c|c|}
  \hline
\multicolumn{2}{|c|}{ State } & Sector & Representation \\
  \hline\hline  
  Tachyon & $t$ & $\theta^1$ & $(\rep1;\rep1,\rep1,\rep2)$ \\
  \hline 
  Blow-up mode & $b$ & $\theta^2$ & $(\rep1;\rep1,\rep2_-,\rep1)$ \\
  \hline 
  Complex scalar &  $c$ & $\theta^3$ & $(\rep1;\rep1,\rep2_-,\rep2)$ \\
  \hline
 \end{tabular}
 \renewcommand{\arraystretch}{1}
\end{center}
On general field theoretical grounds we expect that the effective potential for the tachyon $t$ contains the terms
 \begin{equation} \label{EffPot} 
 V_\text{eff} = - m_t^2\, |t|^2 + |\lambda|^2\, |b|^2 \, |t|^2 + \ldots\,, 
 \end{equation}
where $m_t^2$ is the tachyonic mass. 
Hence, when the blow-up mode takes a sufficiently large VEV, the tachyon becomes massive, assuming that the sign in front of the coupling constant $|\lambda|^2$ is positive. 
This may be motivated as follows:

We argued above that the scalar potential in such Calabi-Yau compactifications looks like in a supersymmetric theory to lowest order approximation. 
If we suppose that this also applies to the effective theory with tachyons, then such coupling should arise from some superpotential. 
Indeed, if the theory was supersymmetric, all orbifold selection rules would allow for a superpotential coupling $\mathcal{W} \supset \lambda TBC$, where the capital letters correspond to hypothetical chiral superfield extensions of the bosons. 
The second term in~\eqref{EffPot} corresponds then to $\left|F_C\right|^2$ where $F_C$ is the auxiliary-field component associated to the superfield containing the complex scalar $c$. 

The crucial point here is that since the blow-up mode $b$ is unique, any full blow-up of this model turns the tachyon into a massive state. 
From the smooth compactification perspective such a coupling comes from a worldsheet instanton.

In this example we have confirmed that the tachyonic state most likely gets decoupled from the low-energy spectrum by a superpotential-like effective scalar potential. 
However, it should be stressed that the dynamics of the tachyon at the orbifold point 
would drive it down the potential to an unknown ground state rather than perform the blow-up. 
Nevertheless, we showed that there is a flat direction in the effective potential connecting 
the orbifold and blow-up theories which explains the disappearance of the tachyon.
Given that on general grounds we know that no tachyons appear on smooth Calabi-Yau manifolds, 
we expect similar mechanisms will be at work for any orbifold of the N=0 theory containing twisted tachyons.

\section{Model searches on Abelian orbifolds}
\label{sc:Models}


In this section, we report on model scans for Calabi-Yau orbifold compactifications of the N=0 SO(16)$\times$SO(16) string. 
The essential objectives of this study are the following: 
\begin{itemize}
\item
To have highly frequent and non-trivial cross-checks on our construction. 
\item 
To obtain some relative estimate on how abundant the tachyonic models are on orbifolds with twists that in principle allow for twisted tachyons. 
\item 
To show that it is possible to obtain tachyon-free Standard Model-like models and give some conservative estimate of how many such models arise on the various orbifold geometries. 
\item 
To obtain some first indications on the issues one has to deal with, to go from Standard-Model-like string models to more realistic constructions. 
\end{itemize} 

\subsection{Automatization of the construction of non-supersymmetric models}


Given that this approach to non-supersymmetric heterotic model building in four dimensions is considered for the first time\footnote{The only other extensive scan~\cite{Dienes:2006ut}
 for non-supersymmetric models in the heterotic context, which we are aware of, used the free fermionic formulation.}, we had to develop new computer codes or heavily modify existing ones, like the ``Orbifolder''~\cite{Nilles:2011aj}. 
In order to have some cross-checks on the results we used three codes to determine orbifold spectra: 
\begin{enumerate}[1.)]  
\item 
A mathematica code that implements the orbifold model construction on the level of partition functions using the fermionic formulation~\eqref{SO(16)_orbifold_partition}.
\item 
A modification of the ``Orbifolder'' code that implements the combination of lattices as dictated by Table~\ref{tb:Lattices_E8_N=0}. 
\item 
A modification of the ``Orbifolder'' code that implements the N=0 theory as a $\Intr_2$ orbifold of the supersymmetric E$_8\times$E$_8$ theory~\eqref{N=0_orbifold_partition}. 
\end{enumerate} 
The modified ``Orbifolder'' codes work for compact toroidal orbifolds with Wilson lines. 
When comparing the results of these different algorithms for the purpose of cross-checks, one should keep the following issues in mind. 
Ignoring the generalized discrete torsion phases, these constructions are not fully equivalent: 
As we saw in eqs.~\eqref{ModInv_ferm_tot} and~\eqref{ModInvWitten}, 
the constraints on the input data differ substantially and certain phases may lead to a complex conjugation of the spectra. 
Hence, only the models with input data that satisfy the stronger constraint $\frac14 e_8^T V_a \equiv 0$ can be directly compared. 
Further important consistency checks on the four dimensional fermionic spectra are the absence of non-Abelian gauge anomalies and universality~\eqref{universalU(1)} of at most a single anomalous U(1). 
These checks are implemented in the ``Orbifolder'' code on the level of chiral superfields~\cite{Nilles:2011aj}. 
Therefore, we modified this in order to consider the actual chiral fermions arising in the N=0 theory taking the opposite ten dimensional 
chiralities of $\Psi_\pm$ (see e.g.\ Table~\ref{tb:N=0Spectrum}) into account. 


Anomaly considerations provide stringent checks on the fermionic spectra, unfortunately such checks do not exist for scalars, hence in particular here having various of procedures to determine the scalar (tachyonic) spectra is very important. 
After a direct comparison of the two implementations 2.) and 3.) of modifications of the ``Orbifolder'' code, 
we cross-checked the results by two independent methods: 
i.) we expanded the full partition functions implemented in the mathematica code 1.) to read off the tachyonic and massless scalar spectrum; 
ii.) as mentioned in Section~\ref{sc:Smooth}, we investigated the resolutions of certain orbifold models using line bundles and compared the bosonic and fermionic spectra before and after the blow-up. 
Even though the various cross-checks mentioned here do not fully ensure that all computed spectra are correct, they certainly ensure that many possible simple or more systematic mistakes have been avoided. 

\subsection{Non-supersymmetric ``Orbifolder'' model scans}

\begin{table}
\begin{center}
\renewcommand{\arraystretch}{1.2}
\scalebox{1}{
%
%
\begin{tabular}{|cc||r|r|r|c|c|}
\hline 
\multicolumn{2}{|c||}{Orbifold} & \multicolumn{1}{c|}{Inequivalent}  & \multicolumn{1}{c|}{Tachyon-free} &
\multicolumn{3}{c|}{SM-like tachyon-free models}
\\ 
\multicolumn{2}{|c||}{~~twist~~  \#(geom)~~} 
& \multicolumn{1}{c|}{scanned models} & \multicolumn{1}{c|}{percentage} & \multicolumn{1}{c|}{~~~~total~~~~} & one-Higgs & two-Higgs\\
\hline\hline
$\Intr_3$ & (1)           &     74,958\phantom{....} & 100\,\%\phantom{......} & 128\phantom{....} & 0 & 0 \\
\hline
$\Intr_4$ & (3)           &  1,100,336\phantom{....} & 100\,\%\phantom{......} & 12\phantom{....} & 0 & 0 \\
\hline
$\Intr_\text{6-I}$ & (2)  &    148,950\phantom{....} & 55\,\%\phantom{......} & 59\phantom{....} & 18 & 0 \\
\hline
$\Intr_\text{6-II}$ & (4) & 15,036,790\phantom{....} & 57\,\%\phantom{......} & 109\phantom{....} & 0 & 1 \\
\hline
$\Intr_\text{8-I}$ & (3)  &  2,751,085\phantom{....} & 51\,\%\phantom{......} & 24\phantom{....} & 0 & 0 \\
\hline
$\Intr_\text{8-II}$ & (2) &  4,397,555\phantom{....} & 71\,\%\phantom{......} & 187\phantom{....} & 1 & 1 \\
\hline
\hline
$\Intr_2 \times \Intr_2$ &(12) &  9,546,081\phantom{....} & 100\,\%\phantom{......} & 1,562\phantom{....} & 0 & 5 \\
\hline
$\Intr_2 \times \Intr_4$ & (10)  & 17,054,154\phantom{....} & 67\,\%\phantom{......} & 7,958\phantom{....} & 0 & 89 \\
\hline
$\Intr_3 \times \Intr_3$ & (5)  & 11,411,739\phantom{....} & 52\,\%\phantom{......} & 284\phantom{....} & 0 & 1 \\
\hline 
$\Intr_4 \times \Intr_4$ & (5)  & 15,361,570\phantom{....} & 64\,\%\phantom{......} & 2,460\phantom{....} & 0 & 6 \\
\hline
\end{tabular}

}
\renewcommand{\arraystretch}{1}
\end{center}
\caption{\label{tb:OrbifoldScanOverview} 
Results of our model search on various $\Intr_N$ and $\Intr_M \times \Intr_N$ orbifold geometries. 
The number of such geometries per orbifold twist is displayed in brackets. 
In the next column we indicate the number of inequivalent models generated, the percentage of them which are non-tachyonic, how many (tachyon-free) Standard Model-like models were found and the frequency among them of models with one or two Higgs scalars. 
}
\end{table} 



To set up a scan for non-supersymmetric models on various Calabi-Yau orbifolds, with orbifold twists listed in the first column of Table~\ref{tb:OrbifoldScanOverview}, we have chosen to work using the implementation 3.) of the list above, because this description has the weakest conditions on the input data of the model. 
A given orbifold twist corresponds to a number of orbifold geometries depending on the number of lattices compatible with this twist; the number of such compatible geometries is indicated in parentheses next to the corresponding twist in Table~\ref{tb:OrbifoldScanOverview}. 
We follow the classification of such Abelian orbifolds as in Ref.~\cite{Fischer:2012qj} which completed the partial classification of Ref.~\cite{string_compactification_phys_rept,RamosSanchez:2008tn}. 
For a given orbifold geometry we randomly generate the input data, i.e.\ shift(s) and Wilson lines, to construct orbifold models. 
We have only collected models which are considered to be inequivalent in the following sense: 
Two orbifold models on the same orbifold geometry are equivalent when they have identical massless bosonic and fermionic and possibly tachyonic spectra up to charges under Abelian factors. 


In Table~\ref{tb:OrbifoldScanOverview} we list the number of inequivalent models we have considered in our scans and indicated which percentage of them is tachyon-free. 
Since our scans have not been systematically exhaustive, we certainly do not wish to imply that the number of inequivalent models will be closely related to the actual figures each of these geometries could actually demonstrate. 
However, we checked that the percentages quoted in this Table do not change significantly when scanning over large sets of models. 
This suggests that these percentages of non-tachyonic models have significant meaning within the limitations of our non-exhaustive scans. 
We see that, when tachyons would be possible on given orbifolds according to Table~\ref{tb:TachyonicOrbifolds}, they arise abundantly yet not predominantly. 


Within the set of non-tachyonic models, we have searched for models which one could call Standard Model-like. 
Our definition of Standard Model-like consists of the following requirements: 
\begin{enumerate}[i.)]  
\item 
The gauge group contains the Standard Model gauge group with the SU(5) normalization of the non-anomalous hypercharge $Y$. 
\item 
There is a net number of three generations of chiral fermions. 
\item 
There is at least one Higgs scalar field.
\item
The exotic fermions are vector-like w.r.t.\ the Standard Model gauge group.
\end{enumerate}
Following this definition we have collected the number of Standard 
Model-like models for each of the scanned orbifolds in the fourth column of Table~\ref{tb:OrbifoldScanOverview}. 
Finally, we have explored the number of Higgs scalar fields found in these semi-realistic models. We list
in the last two columns of Table~\ref{tb:OrbifoldScanOverview} how many of them exhibit one or two Higgs scalars.

\subsection{A tachyon-free Standard Model-like model}
\label{sc:SMsingleHiggs} 

\begin{table}
\begin{center}
\renewcommand{\arraystretch}{1.2}
\scalebox{1}{
\begin{tabular}{|c||c|}
\hline 
Sector & Massless spectrum: chiral fermions / \colorbox{lightgray}{complex bosons}
\\ \hline\hline 
Observable & 
$ 3 (\mathbf{3},  \mathbf{2})_{1/6} 
+ 3 (\overline{\mathbf{3}}, \mathbf{1})_{-2/3} 
+ 6 (\overline{\mathbf{3}}, \mathbf{1})_{1/3} 
+ 3 (\mathbf{3}, \mathbf{1})_{-1/3} 
+ 3 (\mathbf{1}, \mathbf{1})_{1}
+ 5 (\mathbf{1}, \mathbf{2})_{-1/2}  
$
\\ 
& $2 (\mathbf{1}, \mathbf{2})_{1/2} + 20 (\mathbf{1}, \mathbf{1})_{1/2} + 20 (\mathbf{1}, \mathbf{1})_{-1/2} 
+ 6 (\mathbf{3}, \mathbf{1})_{1/6} +  6 (\overline{\mathbf{3}}, \mathbf{1})_{-1/6} 
+ 2(\mathbf{1}, \mathbf{2})_0  
$
\\ \hline 
Obs.\ \& Hid. & 
$3 ( \mathbf{1},  \mathbf{1};  \mathbf{1},  \mathbf{2})_{1/2}
+ 3 ( \mathbf{1},  \mathbf{1};  \mathbf{1},  \mathbf{2})_{-1/2}$
\\ \hline 
Hidden & 
$14 (   \mathbf{1},   \mathbf{2})_0
+10 (\overline{\mathbf{4}},   \mathbf{1})_0
+ 6 (   \mathbf{4},   \mathbf{1})_0
+ 3 (   \mathbf{6},   \mathbf{1})_0
+ 2 (   \mathbf{4},   \mathbf{2})_0
+ 71 (\mathbf{1})_0$
\\ \hline\hline  
\cellcolor{lightgray} Observable & 
\cellcolor{lightgray}$(\mathbf{1}, \mathbf{2})_{-1/2} $ \\
\cellcolor{lightgray}&\cellcolor{lightgray} $ (\mathbf{3}, \mathbf{1})_ {1/6}
+  (\overline{\mathbf{3}},\mathbf{1})_{-1/6} 
+ 2(\overline{\mathbf{3}},  \mathbf{1})_{1/3} 
+13( \mathbf{1},  \mathbf{2})_{0} 
+20( \mathbf{1},  \mathbf{1})_{-1/2} 
+18( \mathbf{1},  \mathbf{1})_{1/2} 
$
\\ \hline 
\cellcolor{lightgray}Obs.\ \& Hid. & \cellcolor{lightgray}
$ (  \mathbf{1},   \mathbf{1};   \mathbf{4},   \mathbf{1})_{1/2} 
+ (   \mathbf{1},   \mathbf{1};   \mathbf{4},   \mathbf{1})_ {-1/2}
+ (   \mathbf{1},   \mathbf{2};   \mathbf{1},   \mathbf{2})_{0}
$
\\ \hline 
\cellcolor{lightgray}Hidden & \cellcolor{lightgray}
$ 14 (   \mathbf{1},   \mathbf{2})_0
+  4 (   \mathbf{4},   \mathbf{1})_0
+    (   \mathbf{6},   \mathbf{2})_0
+ 23 (\mathbf{1})_0 $
\\ \hline 
\end{tabular}}
\renewcommand{\arraystretch}{1}
\end{center}
\caption{\label{tb:OneHiggsModel} 
Spectrum of a non-supersymmetric one-Higgs doublet model derived from string theory.  
The states are divided into fermionic and bosonic classes and we distinguish 
whether they are only charged under the Standard Model group $G_\text{obs}$, the hidden group $G_\text{hid}$ or both. 
}
\end{table}


Let us discuss one specific tachyon-free Standard Model-like theory explicitly.
This is a model defined on the $\Intr_\text{6-I}$ orbifold on the lattice SU(3)$^3$. 
The shift and Wilson lines of this model are given by
\begin{subequations}
\begin{align}
%
%
V &= 
\frac 1{6}\big( 3, -3,  -3,   1,     -4,  -3,   0,   1\big)\big( 3,  -4,  -3,   0,     -3,   0,   1,   4\big)~,
\\[1ex] 
A_{5,6} &= 
\frac 16\big( 5, 3,  3,  -3,    -1,  3,   1,   -3\big)\big( -7,  3,  -1,   -5,    1,   5,  3,  -1\big)~. 
\end{align} 
\end{subequations} 
Up to additional U(1)s, the observable and hidden gauge groups are: 
\begin{equation}
G_\text{obs} = \text{SU(3)}_C \times \text{SU(2)}_L \times \text{U(1)}_Y~, 
\qquad 
G_\text{hid} = \text{SU(4)} \times \text{SU(2)}~. 
\end{equation}
Hence, we have states charged only under $G_\text{obs}$, only under $G_\text{hid}$, and very few exotic states charged under both gauge sectors. 
In Table~\ref{tb:OneHiggsModel} we present the quantum numbers of the full massless spectrum 
for both the fermions and the bosons, indicating the hypercharge as subindex.
In order not to clutter the notation too much, we only indicate the representations w.r.t.\ either the observable / hidden group for those states that are only charged under either one of them, and omit all U(1) charges excepting the hypercharge.
For the states that are charged under both the hidden and observable groups, we use a semicolon to separate their representations. 

It is not difficult to see that this model has precisely a single scalar Higgs $(\mathbf{1},\mathbf{2})_{1/2}$ 
and contains three generations of Standard Model quarks and leptons. 
However, we see that the model also contains many states that should be considered as exotics w.r.t.\ the Standard Model, i.e.\ states charged under the Standard Model gauge group $G_\text{obs}$ while not being part of it.  
Note that this definition differs from the definition of exotics for MSSM-searches in the supersymmetric situation: 
In the latter case, all the scalar superpartners of the Standard Model fermions are not considered to 
be as exotics w.r.t.\ the MSSM although they are exotics w.r.t.\ the Standard Model itself. 
In particular, the model presented here contains complex scalar SU(3)$_C$ triplets, which -- like analogous states in MSSM-like models -- might take VEVs and lead to color-breaking vacua unless there exists a mechanism that forbids such VEVs. 
On the other hand, similarly to what happens in known string MSSM candidates, all exotic fermions of the model are vector-like w.r.t.\ the Standard Model and can thus be decoupled from the emerging field theory in scalar minima, where the scalars $(\boldsymbol1)_0$ coupling to the fermions could develop non-vanishing VEVs.\footnote{A detailed discussion of this issue is beyond the scope of this work and will be carried out elsewhere.}
Hence, we see that this model faces similar phenomenological challenges as its supersymmetric counterparts.

\section{Discussion}
\label{sc:Conclusion}

In this paper we have investigated non-supersymmetric model building within string theory. 
Concretely, we considered compactifications of the non-supersymmetric heterotic SO(16)$\times$SO(16) string on singular orbifolds and smooth Calabi-Yau manifolds which themselves would preserve N=1 supersymmetry in compactifications of N=1 string theories. 

We have reviewed two formulations of the N=0 theory: 
A fermionic formulation with certain generalized discrete torsion phases switched on and a lattice formulation. 
We found that the lattice formulation appears to be more flexible when 
it comes to orbifold compactifications mainly because the conditions for 
modular invariance and consistent projections are less constraining. 
The reason why the conditions are stronger in the fermionic description 
can be traced back to the requirement of obtaining well-defined GSO 
and orbifold projections from these generalized torsion phases.

We have investigated various aspects of orbifold compactifications of the N=0 heterotic theory. 
First of all, twisted tachyonic scalars may appear at tree level depending on both 
i) the geometry of the orbifold and ii) the specifics (shifts and Wilson lines) of each model.
In Table~\ref{tb:TachyonicOrbifolds} we indicate for which orbifold geometries 
tachyonic states may arise and for which this is impossible. 
In addition, we have checked that the orbifold compactification spectra of this N=0 
theory are always free of non-Abelian anomalies and that there is at most  one
anomalous U(1) satisfying the conventional universality conditions~\eqref{universalU(1)}. 

We also considered smooth Calabi-Yau compactifications of the non-supersymmetric heterotic SO(16)$\times$SO(16) theory. 
The fermionic spectra can be computed using the usual index theorems or cohomology methods. 
In addition, exploiting that the compactification manifold is Calabi-Yau, we realized that the spectrum 
of scalar bosons is also dictated by the zero modes of the Dirac operator of the -- non-existing -- 
superpartners of the gauge fields. 
We used this to perform a cross-check of the computation of the scalar spectra on the orbifold: 
We showed that the bosonic spectra on the orbifold and smooth resolution agreed up to 
vector-like states that decouple when the blow-up modes attain non-trivial VEVs. 
This showed that on smooth manifolds the N=0 theory never leads to tachyons when the large volume approximation applies. 
Consequently, when one fully blows up an orbifold, all the tachyons should disappear from the spectrum. 
To illustrate these features, we considered in Subsection~\ref{sec:blowup_tachyon_model} the blow-up of a tachyonic $T^6/\Intr_\text{6-I}$ model, where all tachyonic contributions can be decoupled 
by considering superpotential-like scalar interactions. 

Finally, we have performed model searches on orbifolds of the non-supersymmetric heterotic theory. 
With this purpose, we have developed three independent codes to determine spectra of such models.
Two of them are modifications of the publicly available ``Orbifolder'' package while the other 
is a Mathematica code which implements the partition functions as a whole. 
We confirmed that the results of these different codes agree when the input data is suitably chosen. 
We used the modified ``Orbifolder'' based on the construction of the non-supersymmetric 
SO(16)$\times$SO(16) theory as a (freely-acting) orbifold of the E$_8\times$E$_8$ theory, 
to build more than $76$ million inequivalent consistent four-dimensional N=0 string orbifold models with different geometries. 
By means of this extensive model search, we showed that for orbifolds on which tachyons are not strictly forbidden, they appear abundantly but not predominantly. 
In addition, we found that it is possible to generate Standard-Model-like theories with a net 
number of three generations of chiral Standard Model fermions and at least one Higgs scalar. 
As can be seen from Table~\ref{tb:OrbifoldScanOverview}, we constructed over 12,000 models of this type.  
One of these models with one Higgs field is briefly discussed in Subsection~\ref{sc:SMsingleHiggs}.

\subsection*{Outlook}

Let us conclude with some remarks about known challenges that the approach presented in this work faces:  


The most notorious and difficult problem is that of the cosmological constant and the related dilaton tadpole. 
This means that the theory is unstable since the dilaton is driven away from its perturbative value 
in which the analysis of the model was performed. 
This might lead to interesting but problematic properties in the cosmological setting~\cite{Dudas:2000ff,Dudas:2002dg}. 
In supersymmetric string theories the cosmological constant is identically zero since no tachyons are present and at each massless level one encounters an equal number of bosonic and fermionic states.
In the orbifold compactifications of the non-supersymmetric SO(16)$\times$SO(16) theory we have seen that, except for $\Intr_3, \Intr_4$ and $\Intr_2\times\Intr_2$ orbifolds, one is not safeguarded from tachyons in the spectrum.  
When they occur, then the computation of the cosmological constant leads to a divergent result. 
However, even if we restrict to the orbifolds which are always tachyon-free, or consider only the tachyon-free models for the orbifolds that might possess tachyons, then all states at all mass level contribute to the cosmological constant. 
Moreover, as emphasized in \cite{Dienes:2006ut} even non-level-matched states contribute to the cosmological constant. 
The off-shell tachyonic states which only exist in the loop give rise to the most sizable contributions of all non-level-matched states. 
All in all this means that even though string theory does give a finite result of the cosmological constant for non-tachyonic models, this result will generically be large. 
Moreover, it will be very sensitive to the values of all kind of moduli like the torus radii or size of blow-up modes. 
There have been some attempts in the literature to address the issue of the vacuum energy in such string theory context: 
In Ref.~\cite{Kachru:1998hd} a certain non-supersymmetric non-Abelian orbifold of the type-II string was considered. 
Moore has considered an Atkin-Lehner symmetry to enforce to have a vanishing partition function without target space supersymmetry~\cite{Moore:1987ue} (for an extension see~\cite{Dienes:1990qh}).


Similar issues one expect for the computation of the Higgs mass in these non-supersymmetric models.
To determine the Higgs mass one should compute the two-point function of the vertex operators corresponding to the Higgs field and its conjugate. 
On general grounds one again expects that the result is finite, but generically the value of the Higgs mass will be of the order of the string scale unless one could somehow impose some very non-trivial cancellations. 


Both these problems are well-known problems within the Standard Model and beyond. 
The only real candidate to address these issues is supersymmetry. 
However, even if one assumes that supersymmetry is broken at a low scale compared to the string scale, such that current experimental data suggest a moderate hierarchy between the Higgs mass and the supersymmetry breaking scale, then still this will lead generically to a huge cosmological constant as compared to the observed value. 
Hence, the non-supersymmetric models considered here or in ref.~\cite{Dienes:2006ut} are 
at least at equal footing as generic field theories that incorporate the Standard Model. 
One might even argue that the string constructions are more under control than such effective 
field theories as the non-supersymmetric string models at least allow one to definitely 
calculate the Higgs mass and the cosmological constant in principle. 


In this work we have considered supersymmetric backgrounds for a non-supersymmetric theory 
and seen that many consequences of supersymmetry still hold at tree level and leading order in $\alpha'$. 
It would therefore be very interesting to investigate how non-supersymmetric features arise at higher orders and non-perturbatively. 
This is very important as it is expected that such effects may (re)introduce tachyons in the description. 
Moreover, it would be interesting to see how the supersymmetric backgrounds we considered get corrected by quantum effects. 


In addition, the non-supersymmetric models constructed in this work will face similar questions as their supersymmetric counter parts. 
As our Standard Model-like single Higgs model discussed in Subsection~\ref{sc:SMsingleHiggs} illustrates, such models will suffer from having many additional exotic states. 
Like in the supersymmetric case, one might try to find vacuum configurations where all these exotics decouple.

\subsection*{Acknowledgements}

We would like to thank Brent Nelson for a discussion at ICTP which initiated this project. 
We are indebted to Panagiotis Athanasopoulos, Emilian Dudas, Keith Dienes, Ioannis Florakis, Andre Lukas and Viraf Mehta for very valuable discussions. 
We especially would like to thank Patrick Vaudrevange for very helpful discussion on how to implement the non-supersymmetric theory in the ``Orbifolder'' package. 

SGN thanks ICTP, Trieste, and MB the LMU University, Munich, for kind hospitality. 
SGN and MB are very grateful for their stay at the Department of Theoretical Physics of the UNAM University in Mexico. 
Finally, SRS acknowledges the ICTP, LMU and TUM Universities, Munich, for hospitality during the completion of this work. 

This work was supported by the LMUExcellent Programme, 
by CONACyT grant 151234 and DGAPA-PAPIIT grant
IB101012-RR181012,
and the {\it Cluster of Excellence `Precision Physics, Fundamental Interactions and Structure of Matter' (PRISMA)} DGF no. EXC 1098.
OL acknowledges the support by the DAAD Scholarship Programme ``Vollstipendium f\"{u}r Absolventen von deutschen Auslandsschulen'' within the ``PASCH--Initiative''.

\appendix 
\def\theequation{\thesection.\arabic{equation}} 
\setcounter{equation}{0}

\section{Theta functions and lattice sums}
\label{sc:ThetaLattice}

\subsection{Basic partition functions}

The eta and genus-$d$ theta functions are defined as
\begin{equation}
\eta(\tau) = q^{1/24} \prod_{n=1}^\infty \left( 1 - q^n \right) \,, 
\qquad 
\theta_d \brkt{a}{b }(\tau) = \sum_{n \in \mathbb{Z}^k} q^{\frac12(n-a)^2} e^{-2\pi i b^\text{T} \left(n-a\right) } \,,
\end{equation}
with $q = e^{2\pi i\tau}$ in terms of the Teichmm\"uller parameter $\tau$. 

For a set of $d$ complex worldsheet Fermions $\psi^i$ with boundary conditions 
\begin{equation} 
\psi^i(\sigma+1) = e^{2\pi i \alpha_i} \, \psi^i(\sigma)~, 
\qquad 
\psi^i(\sigma-\tau) = e^{2\pi i \beta_i} \, \psi^i(\sigma)~, 
\end{equation}
we find the partition function
\begin{equation} \label{fermion_partition} 
{\mathbf Z}_d\left[^{\alpha}_{\beta }\right](\tau)  = 
\frac{\theta_d\brkt{\frac 12 e_d - \alpha}{\frac 12 e_d - \beta }(\tau)}{\eta^d(\tau)}~,  
\end{equation} 
where $e_d = (1,\ldots, 1)$ is the $d$ component vector with only entries equal to $1$. By including appropriate phases,  
\begin{equation} \label{sector_phase}
\widehat {\mathbf Z}_d\brkt{\alpha}{\beta}(\tau) 
= e^{- \pi i \alpha^T(\beta-e_d)}\, {\mathbf Z}_d\brkt{\alpha}{\beta}(\tau)~,
\end{equation} 
one can ensure that this partition function is modular invariant up to phases that only depend on $d$: 
\begin{equation} \label{modular_trafo_phases}
\widehat {\mathbf Z}_d\brkt{\alpha}{\beta}(\tau+1) 
=  e^{2\pi i \frac d{12}}\,  
\widehat {\mathbf Z}_d\brkt{\alpha}{\beta+\alpha}(\tau)~, 
\qquad 
\widehat {\mathbf Z}_d\brkt{\alpha}{\beta}(\tfrac {\text{-}1}\tau) 
= e^{-2\pi i \frac d4}\,\widehat {\mathbf Z}_d\brkt{\beta}{\text{-}\alpha}(\tau)~. 
\end{equation} 
Notice that the additional phase drops out when one has multiple of eight complex fermions with $\Intr_2$ twisted boundary conditions while for four complex fermions one has relative sign: 
\begin{equation} 
\widehat {\mathbf Z}_8\brkt{\frac t2 e_8}{\frac{t'}2 e_8 }(\tau) =
{\mathbf Z}_8\brkt{\frac t2 e_8}{\frac{t'}2 e_8 }(\tau)~, 
\qquad 
\widehat {\mathbf Z}_4\brkt{\frac s2 e_4}{\frac{s'}2 e_4 }(\tau) =
(-)^{s's} 
{\mathbf Z}_4\brkt{\frac s2 e_4}{\frac{s'}2 e_4 }(\tau)~. 
\end{equation}

For a set of $d$ complex bosons $X^i$ with identical boundary conditions we find instead 
\begin{equation}  \label{twisted_boson_partition} 
{\mathbf Z}_d^X\brkt{\alpha}{\beta}(\tau,\overline\tau) = 
\frac{1}{\dsp \left| {\mathbf Z}_d\brkt{\alpha}{\beta}(\tau)\right|^2}~. 
\end{equation}
When $D$ real bosons $x^\mu$ have trivial twist boundary conditions their partition function becomes 
\begin{equation} \label{free_boson_partition} 
{\mathbf Z}_D^x(\tau,\overline\tau) = \frac1{\tau_2^{D/2} | \eta(\tau)|^{2D}}~. 
\end{equation} 

\subsection{Lattice sums}

\begin{table}
 \begin{center}
  \renewcommand{\arraystretch}{1.2}
  \begin{tabular}{|c||c|c|}
   \hline
   &Weight lattice & Lattice vectors \\
   \hline\hline 
   {\bf R}$_D$  & Root & 
   $n \in \mathbb{Z}^D$, $\sum n_i \in 2\mathbb{Z}$ 
    \\ \hline
   {\bf V}$_D$ & Vector & 
   $n \in \mathbb{Z}^D$, $\sum n_i \in 2\mathbb{Z}+1$ 
   \\ \hline
   {\bf S}$_D$ & Spinor & 
   $n \in \mathbb{Z}^D + \frac 12 e_D$, $\sum n_i \in 2\mathbb{Z}$ 
    \\ \hline
   {\bf C}$_D$ & Cospinor  & 
   $n \in \mathbb{Z}^D + \frac 12 e_D$, $\sum n_i \in 2\mathbb{Z}+1$ 
   \\ \hline \hline 
  $ \boldsymbol{\Gamma}_4$ & Space-time & $\text{\bf V}_4 \oplus \text{\bf S}_4$
   \\ \hline 
   {\bf E}$_8$ & E$_8$ Root & $\text{\bf R}_8 \oplus \text{\bf S}_8$
   \\ \hline 
   $\boldsymbol{\Gamma}_{16}$ & E$_8\times$E$_8$ Root &  $\text{\bf E}_8 \oplus \text{\bf E}_8$
   \\ \hline 
     \end{tabular}
       \renewcommand{\arraystretch}{1}
 \end{center}
\caption{ \label{tb:Lattices} 
Definition of various weight lattices.}
\end{table}

Let $\boldsymbol{\Lambda}_D$ be $D$ dimensional real lattice on which left-moving bosons $Y^I$ live. They lead to the partition function 
\begin{equation}
\boldsymbol{\Lambda}_D(\tau) = \frac 1{\eta^{D}}\, 
\sum_{P\in \gL_D}\, q^{\frac 12 P^2}~. 
\end{equation} 
To resolve the slight abuse of notation using the same notation for a lattice and its associated partition function we indicate the partition functions by always giving its $\tau$ argument. 
The partition functions for the lattices defined in Table~\ref{tb:Lattices} can be written as: 
\begin{subequations}
 \begin{align}
{\text{\bf S}_D/\text{\bf C}_D}(\tau) = 
\frac12 \left( {\mathbf Z}_D\brkt{0}{e_D/2}(\tau) \pm {\mathbf Z}_D\brkt{0}{0}(\tau)  \right) 
&= q^{D/12} \left( 2^{D-1}  +\mathcal{O}(q) \right)~,  
\\ 
{\text{\bf R}_D}(\tau) = 
\frac12 \left({\mathbf Z}_D\brkt{e_D/2}{e_D/2}(\tau) + {\mathbf Z}_D\brkt{e_D/2}{0}(\tau) \right) 
&= q^{-D/24} \left( 1 + D(2D-1) q +\mathcal{O}(q^2) \right)~, 
\\ 
{\text{\bf V}_D}(\tau) = 
\frac12 \left( {\mathbf Z}_D\brkt{e_D/2}{e_D/2}(\tau)  - {\mathbf Z}_D\brkt{e_D/2}{0}(\tau) \right) 
&= q^{-D/24 + 1/2} \left( 2D + \mathcal{O}(q) \right)~. 
 \end{align}
 \end{subequations} 
The partition function associated with the lattice $\boldsymbol{\Gamma}_4$ also encodes the target space spin-statistics:
\begin{equation}
{\boldsymbol{\Gamma}_4}(\tau)  
= \frac 1{2\eta^4} 
\sum_{k,k'=0}^1
\sum_{n \in \Intr^4} q^{\frac 12 (n+\frac k2 e_4)^2} \, 
e^{2\pi i \frac {k'}2 e_4^T n} \, 
(-)^{k'k + k' + k}~, 
\end{equation} 
since 
\begin{equation} \label{Space-time_partition} 
{\boldsymbol{\Gamma}_4}(\tau) = {\text{\bf V}_4}(\tau) - {\text{\bf S}_4}(\tau)  
= - \frac 12\, \sum_{s,s'=0}^1 \widehat {\mathbf Z}_4\brkt{\frac s2 e_4}{\frac{s'}2 e_4}(\tau)~. 
\end{equation}
Similarly, we have 
\begin{equation} \label{E8_partition} 
{\text{\bf E}_8}(\tau) = {\text{\bf R}_4}(\tau) + {\text{\bf S}_8}(\tau) 
= \frac 12\, \sum_{t,t'=0}^1 \widehat {\mathbf Z}_8\brkt{\frac t2 e_8}{\frac{t'}2 e_8}(\tau)~,
\qquad 
\mathbf{\Gamma}_{16}(\tau) = {\text{\bf E}_8}(\tau) \cdot {\text{\bf E}_8}(\tau)~.   
\end{equation}

Furthermore, to describe orbifolded lattices sums, we define the shifted lattice partition function  
\begin{equation} \label{lattice_partition_orbifolded} 
\widehat{\mathbf{\gL}}_D \brkt{\alpha}{\beta }(\tau)   
= \frac 1{\eta^D} 
\sum_{P\in \gL_D} q^{\frac 12 (P+\alpha)^2} \, 
e^{2\pi i \beta^T P} \, 
e^{\pi i\, \alpha^T \beta}~
\end{equation} 
associated to any lattice $\Lambda_D$. 
When $\Lambda_D$ is a direct sum of two lattices, then this definition is taken to linear in the sense that e.g.
\begin{equation}
\widehat{\boldsymbol{\Gamma}}_4 \brkt{\alpha}{\beta }(\tau) = 
\widehat{\text{\bf V}}_4 \brkt{\alpha}{\beta }(\tau) 
- \widehat{\text{\bf S}}_4 \brkt{\alpha}{\beta }(\tau)~. 
\end{equation} 
The shifted lattice sum~\eqref{lattice_partition_orbifolded} already includes appropriate vacuum phases such that for the Euclidean lattices we have 
\begin{equation} \label{lattice_modular_trafo_phases}
\widehat {\mathbf{\gL}}_D\brkt{\alpha}{\beta}(\tau+1) 
=  e^{-2\pi i \frac D{24}}\,  
\widehat{\mathbf{\gL}}_D\brkt{\alpha}{\beta+\alpha}(\tau)~, 
\qquad 
\widehat{\mathbf{\gL}}_D\brkt{\alpha}{\beta}(\tfrac {\text{-}1}\tau) 
= \widehat{\mathbf{\gL}}_D\brkt{\beta}{\text{-}\alpha}(\tau)~. 
\end{equation} 
For the lattice partition function ${\boldsymbol{\Gamma}_4}$ we have instead
\begin{equation} 
\widehat{\boldsymbol{\Gamma}}_4\brkt{\alpha}{\beta }(\tau+1) 
= e^{2\pi i/3}\, 
\widehat{\boldsymbol{\Gamma}}_4 \brkt{\alpha}{\beta }(\tau)~,  
\qquad 
\widehat{\boldsymbol{\Gamma}}_4 \brkt{\alpha}{\beta } (\tfrac {\text{-}1}\tau) 
= \widehat{\boldsymbol{\Gamma}}_4 \brkt{\beta}{\sm\alpha} (\tau)~. 
\end{equation} 
\bibliographystyle{paper}
{\small
\bibliography{paper}
}
\end{document}